\documentclass[utf8]{FrontiersinHarvard} 

\usepackage{url,hyperref,lineno,microtype,subcaption}
\usepackage[onehalfspacing]{setspace}
\usepackage{gensymb}
\usepackage{mhchem}
\usepackage[flushleft]{threeparttable}

\newcommand{\solarmass}{~M$_{\odot}$}
\newcommand{\solarrad}{~R$_{\odot}$}
\newcommand{\solarlum}{~L$_{\odot}$}
\newcommand{\gcc}{~g~cm$^{-3}$}
\newcommand{\edit}{}

\def\keyFont{\fontsize{8}{11}\helveticabold }
\def\firstAuthorLast{Young} 
\def\Authors{Alison K. Young\,$^{1,2}$}
\def\Address{$^{1}$SUPA, Institute for Astronomy, University of Edinburgh, Blackford Hill, Edinburgh EH9 3HJ, UK\\
$^{2}$Centre for Exoplanet Science, University of Edinburgh, Edinburgh, EH9 3HJ, UK}
\def\corrAuthor{Alison K. Young}

\def\corrEmail{alison.young@ed.ac.uk}

\begin{document}
\onecolumn
\firstpage{1}

\title[First and second cores]{Insights into the first and second hydrostatic core stages from numerical simulations} 

\author[\firstAuthorLast ]{\Authors} 
\address{} 
\correspondance{} 

\extraAuth{}

\maketitle

\begin{abstract}

\section{}
{\edit

The theory of how low mass stars form from the collapse of a dense molecular cloud core has been well-established for decades. Thanks to significant progress in computing and numerical modelling, more physical models have been developed and a wider parameter space explored to understand the early stages of star formation more fully. In this review, I describe the expected physical properties of the first and second core stages and how the inclusion of different physics affects those predicted characteristics. I provide an overview of chemical models and synthetic observations, looking towards the positive identification of the first core in nature, which remains elusive. However, there are a few likely candidate first cores, which are listed, and I briefly discuss the recent progress in characterising the youngest protostellar sources. Chemistry will be instrumental in the firm identification of the first core so we require robust theoretical predictions of the chemical evolution of protostellar cores, especially of the first and second core outflows. Looking ahead, simulations can shed light on how the protostellar collapse phase shapes the evolution of the protostellar disc. Simulations of dust evolution during protostellar core collapse show there is significant enhancement in grain size and abundance towards the centre of the core. Chemical models show that the warm, dense conditions of the first core drive chemical evolution. There is a wide scope for further study of the role that the first and second core stages play in determining the structure and composition of the protostellar disc and envelope and, of course, the eventual influence on the formation of planets.}

\tiny
 \keyFont{ \section{Keywords:} star formation, hydrodynamics, radiative transfer, magnetohydrodynamics} 
\end{abstract}

\section{Introduction}
{\edit Stars form in collapsing dense cores within giant molecular clouds, a process involving contraction and density increases of several orders of magnitude.}
The concept of a temporary stable phase at the centre of a collapsing protostellar core was introduced in the seminal work of Richard Larson \citep{larson1969}. The results of those first numerical solutions to the equations governing the collapse of a molecular cloud showed that the collapse is extremely non-homologous. Our understanding of the process remains essentially unchanged since it was first described over 50 years ago. Nonetheless, the technological progress of the last few decades, as well as new computational methods, has enabled sophisticated simulations to be performed in 2- and 3-dimensions and with additional physics. The purpose of this review is firstly to summarise the developments that have taken place since the protostellar collapse process was first described; and secondly to demonstrate that there is some variation in the predicted characteristics of the first and second hydrostatic cores
\footnote{A note on definitions.
The first hydrostatic core is also known in other works as the `first core', `first protostellar core', 'first Larson core', `pre-stellar disc', and even as a `Class $-1$ object' in \citet{boss1995}. 
The stable object that forms within the first hydrostatic core after the second collapse is known as the `stellar core', `second hydrostatic core', 'second Larson core', or just `protostellar core'. \citet{boss1989nov} further suggested that the two objects be known as the inner and outer cores, in acknowledgement that the first core persists while the second core forms at its centre. Here we refer to the first and second hydrostatic cores (first and second core for short) and suggest that these descriptive terms are clearest and most informative. In this review, I will use the term `protostellar core' to refer to the molecular cloud core for which collapse is modelled.}. In section \ref{sec:discussion}, I will also discuss what models are hinting at as to the role of the protostellar collapse stages in shaping the properties of protostellar discs and envelopes. Finally, we look to the future at the outlook for comparisons with observations and future directions for numerical simulations of protostellar core collapse.

\subsection{Overview}

Giant molecular clouds are thought to fragment to form protostellar cloud cores, which may then become unstable to gravitational collapse if there is insufficient thermal support.
The now familiar process of an isolated spherical gas cloud collapsing under self-gravity has the following main stages:
\begin{enumerate}
    \item Isothermal collapse\footnote{\edit The progression of the isothermal collapse phase depends on the initial structure of the cloud core. Whether the contraction of the core is best described by the Larson-Penston flow \citep{larson1969,penston1969}, `inside-out' collapse of a singular isothermal sphere \citep{shu1977}, quasi-static collapse of a Bonnor-Ebert sphere \citep[e.g.][]{keto2010aa} or something else continues to be debated, though observations indicate that prestellar cores are supported by thermal rather than magnetic pressure which simplifies the problem somewhat. See e.g. \citet{whitworth1996,ppVdensecores,keto2015} for further details and discussion.}. At first the density and opacity are low so thermal energy can be freely radiated away. The gas is approximately in free-fall.

    \item Formation of the first core. The density rises in the centre of the protostellar core and this region becomes opaque to infrared radiation. The collapse continues adiabatically in the central region and the pressure and temperature rise rapidly. The collapse pauses once hydrostatic equilibrium is reached and a pressure supported body, the first core, is formed. The central density of the protostellar core is roughly $10^{-12}$~g~cm$^{-3}$ when this occurs.

    \item Growth of the first core. The envelope material falls onto the first core, increasing its mass and temperature.
    
    \item Formation of the second core. At temperatures of $\sim 2000$~K hydrogen molecules begin to dissociate. This endothermic process changes the equation of state such that the temperature rise is slowed. The pressure no longer rises fast enough to maintain hydrostatic equilibrium and a second collapse ensues. Very rapidly (within the time scale of years) a new equilibrium is reached with central temperatures of order $10^4$~K and the second core is formed. 
    
    \item The second core accretes the remnants of the first core and the envelope continues to fall in. The second core continues to contract and heat as a pre-main sequence star.
    
\end{enumerate}

\section{Developments since 1969}

A number of assumptions go into modelling protostellar collapse and these have been revisited over the years as new methods have been developed. Here, I will outline the most important developments that have been made since Larson's first description of the gravitational collapse of a protostellar core. In section \ref{sec:radtrans} we look at the difference in outcome of employing grey radiative transfer rather than a barotropic equation of state and also the results of testing frequency dependent radiative transfer and an equation of state updated from live chemistry calculations. In sections \ref{sec:dimensions} and \ref{sec:Bfields} I summarise the progress enabled by 2- and 3-dimensional models and by the inclusion of magnetic fields in the calculations respectively.

\subsection{Radiative transfer}
\label{sec:radtrans}
The evolution through collapse to stellar core formation is driven by changes in opacity and the equation of state. \citet{larson1969} evolved the energy equation and a radiative diffusion equation in his original 1-D calculations, using pre-calculated values of the Rosseland mean opacity over the relevant temperature range. This approach, as he acknowledged, is valid only for optically thick gas and resulted in optically thin gas being set to the boundary temperature (10~K). For that reason, those calculations missed the heating of the inner envelope.

Rather than calculating radiative energy transport, earlier models with 2 and 3 dimensions made the simplifying assumption of a barotropic equation of state \citep[e.g.][]{bate1998,tomisaka2002aa} whereby the gas pressure is given by: 

\begin{equation}
    p = K \rho ^\gamma.
\end{equation}

\noindent Here the exponent $\gamma$ is the ratio of specific heats and $K$ is a constant specific to each value of $\gamma$. To parameterize the effects of the changing opacity and hydrogen dissociation $\gamma$ is prescribed. For the isothermal phase, $\gamma=1$; for the adiabatic phase $\gamma=7/5$; to mimic the dissociation of \ce{H2}, $\gamma\approx1.1$; and for the monatomic gas in the second core $\gamma = 5/3$. The application of a barotropic equation of state captures the qualitative behaviour of the collapse well but radiative transfer is required to reproduce the thermal evolution more accurately \citep{tomida2010may,tomida2010dec,commercon2010aa, bate2011}. The kinetic energy of the infalling material is radiated away at the first core accretion shock and barotropic treatments cannot reproduce this cooling mechanism \citep{commercon2011fc}.

The problem of handling radiative transport in both optically thin and optically thick regions is usually addressed by implementing the flux-limited diffusion (FLD) approximation \citep{levermore1981aa}. An early application of FLD to 2-D models showed that the temperature structure was more spherical than the flattened density structure but was unable to follow the evolution to the formation of the first core due to computational limitations \citep{bodenheimer1990}. Implementing radiative transfer via FLD in 3-D simulations led to higher temperatures than with a barotropic equation of state \citep{whitehouse2006,tomida2010may,bate2011}. The higher temperatures provide increased thermal support which increases the lifetime of the first core. A striking difference was seen using the two-temperature (gas + radiation) model of \citet{whitehouse2006} whereby the lifetime of the first core increased from a maximum of 1500 years with the barotropic equation of state to a maximum of 3000 years \citep{bate2011}. Regarding the second core phase, the inclusion of shock heating increases the second core radius and allows a transient expansion phase \citep{tomida2013aa}. With radiative transfer, the energy released by second core formation transfers rapidly to the surrounding remnants of the first core, launching a bipolar outflow \citep{bate2011}. The shock may expel the first core out to 500~au and this may be a cyclic process \citep{bate2010aa,schonke2011}.

The grey Rosseland mean opacity is normally used but more complex methods have been tested in radiation hydrodynamics models. 1-D simulations have compared the results of modelling the first collapse with frequency--dependent radiative transfer by binning wavelengths into 5 and 10 groups \citep{vaytet2012}. This showed that the radius, mass and temperature of the first core are slightly underestimated with grey radiative transfer but there are no major differences in the outcomes, in agreement with prior results of \citep{masunaga1998}. Continuing to the formation of the second core, the evolution of the first core was similar with frequency-dependent models to the grey models \citep{vaytet2013}.

{\edit Chemical models can be combined with radiation hydrodynamical models to calculate the time-dependent mean molecular mass fractions for the changing density and temperature. The inclusion of basic chemical reactions of the most abundant species in hydrodynamical models showed that the dust evaporation phase can change the structure of the first core \citep{tscharnuter2007,schonke2011}. Dust free optically thin regions provide a pathway for cooling, but an optically thicker region at the edge of the first core can cause it to swell rapidly.
Another approach to handling radiation in the optically thick envelope is to implement a radiative cooling approximation. The optical depth to a parcel of gas is approximated by assuming that it is located within a stable Bonnor-Ebert sphere with a known density profile \citep{stamatellos2007,lombardi2015}. This approach achieves similar results for protostellar collapse to 1-D radiative transfer models at a reduced computational expense.}

In summary, radiative transfer does not change the qualitative picture of protostellar collapse. It does, however, slightly change the thermal evolution and increases the first core lifetime, which makes the first core more likely to be observed.

\subsection{2- and 3- dimensions}
\label{sec:dimensions}
The extension of models of cloud collapse to 2- and 3- dimensions allows for non-spherical structures to be studied which has arguably had the largest impact on study of the first and second cores. The first two-dimensional calculations were able to follow the collapse to the formation of the second core \citep{tscharnuter1987}. The application of a nested grid in axisymmetric calculations allowed disc formation to be resolved in the same calculation as the whole cloud collapse \citep{yorke1993}. These showed that rotation gave rise to a flattened first core. Three-dimensional calculations showed rotational instabilities developed in the first core \citep{boss1989nov}. The bar-mode instability can lead to the formation of a disc which may then fragment \citep[][and see section~\ref{sec:morphology}]{bate1998,bate2011}. In three dimensions, discs, fragments, outflows and jets can form and will be discussed in detail later.

\subsection{Magnetic fields}
\label{sec:Bfields}
The inclusion of magnetic fields allowed for the formation of outflows and a magnetically-supported pseudo-disc \citep[e.g.][]{galli1993,tomisaka2002aa,matsumoto2004,banerjee2006aa,hennebelle2008aa}. However, under ideal magnetohydrodynamics (MHD) angular momentum is extracted very efficiently from the central regions of the core, suppressing the formation of a rotationally-supported disc in what is known as the `magnetic braking catastrophe' \citep{allen2003aa,mellon2008aa,hennebelle2008aa}. This effect is reduced once non-ideal MHD processes are considered but the resulting discs are still smaller than those produced by purely radiation hydrodynamics simulations \citep[e.g.][]{tomida2013aa}. For further discussion of recent results regarding the influence of magnetic fields on protostellar collapse, disc formation and outflow structures see \citet{maury2022}.

\section{Predicted properties from models}

\subsection{Features of first and second core stages} 

\begin{figure}
    \centering
    \includegraphics[width=8cm]{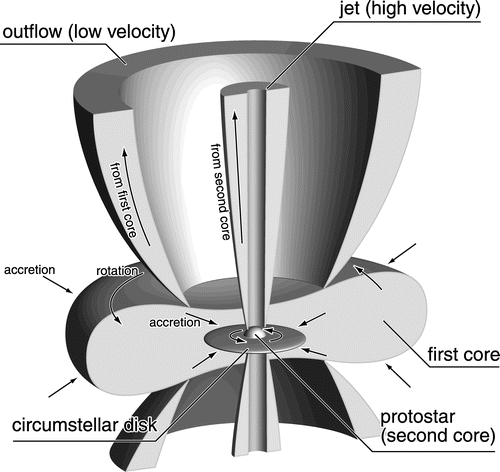}
    \caption{Schematic of the main structures associated with the first and second cores. The first core is flattened in a rotating protostellar core and launches a slow, broad outflow. The second core forms at the centre of the former first core and an accretion disc forms around it. The second core launches a faster, collimated outflow. From \citet{machida2008aa}, Fig 15.\copyright AAS.}
    \label{fig:schematic}
\end{figure}
We saw above that the improvements and extensions to modelling protostellar collapse do not change the main results. The first core forms at the centre of the collapsing cloud core so is embedded within a dense envelope, the properties of which affect its evolution. The boundary of the first core is continuous in density with the infalling envelope, which means that the extent of the first core is not always well defined. \citet{masunaga1998}, for example, suggest the best definition of first core radius is where the gas pressure balances the ram pressure of the infalling envelope. This is where the shock will form at the boundary. Indeed, simulations that resolve that shock use the shock location to define the boundary \citep[e.g.][]{bhandare2018aa}.

Rotation gives rise to discs and MHD models show that an outflow may be launched from both the first and second core. The second core outflow is sometimes referred to as a jet but it is considerably slower than protostellar jets, so we will stick to the term `outflow'. These features are shown in the schematic in Fig.~\ref{fig:schematic}. In this section, I will summarise the predicted properties of the cores, discs and outflows, and which effects are responsible for any variation. A summary of the properties is given for the first core in Table~\ref{tab:FHSCproperties}.

\subsubsection{Masses}

The initial mass of the first and second core when formed are the values typically quoted. Of course, their masses increase substantially with ongoing accretion and the mass a first core reaches before undergoing second collapse may vary considerably depending on rotation, infall and the optical depth of the envelope.

The first estimate of the initial first core mass was 0.01\solarmass, considering a spherically symmetric, non-rotating protostellar core \citep{larson1969}. Since then the masses, radii (in 1-D) and temperatures of first cores have been shown to be largely universal and unaffected by the initial mass or radius of the collapsing protostellar core \citep{masunaga1998,vaytet2012,vaytet2013}. For the initial protostellar core mass range $0.2 \leq  M_{\rm {cc}} \leq 8$\solarmass, 1-D models with frequency--dependent radiative transfer gave rise to first cores with $M_{\rm{fc}} \approx 0.04$\solarmass and  $r_{\rm{fc}} \approx 7$~au \citep{vaytet2017}. 
The thermal evolution of the collapse phase changes slightly when the initial cloud properties are changed. This is due to the corresponding differences in opacity, which determines the rate of cooling and therefore the speed of collapse. In a very cold, and therefore especially unstable, model protostellar core, no first core may form; the collapse proceeds straight to the formation of the second core (\citealt{vaytet2017}; and see also \citealt{stamer2018aa}). Additionally, the radius of the first core is smaller for protostellar cores that are initially more unstable.

Extending the initial mass range to $0.1<M_{\rm{cc}}<100$\solarmass, \citet{bhandare2018aa} demonstrated with 1-D models that there is a dependence of first and second core properties on the initial cloud core mass. Higher mass cores gave rise to larger and more massive first cores up to $\approx$~10\solarmass. For progressively more massive protostellar cores, the first core became smaller and less massive because the ram pressure of infalling material is significant and leads to a high accretion rate. Since the ram pressure is always greater than the thermal pressure for $M\geq40$\solarmass, there is no real first core phase. The central temperature rises rapidly such that hydrogen dissociation occurs and collapse only stops in the centre when the second core forms.

Rotation provides additional support against gravitational collapse, alongside the thermal pressure, which allows the first core to reach a higher mass before the onset of second collapse. Rotation is described by the quantity $\beta = E_{\rm{rot}}/E_{\rm{grav}}$, the ratio of rotational energy to gravitational potential energy. With a barotropic equation of state, the first core mass increased from 0.01\solarmass~ to a maximum of 0.1\solarmass~ with rotation \citep{saigo2008aa}. In RHD simulations, the first core mass increased from $\approx 0.005$\solarmass~ for $\beta=0$ to $\approx 0.22$\solarmass~ for $\beta=0.01$ before the onset of second collapse \citep{bate2011}. With magnetic fields, the first core mass is in the range 0.02-0.05\solarmass~ \citep{commercon2011fc,bate2014}

The initial mass of the second core is $\approx0.0015$\solarmass \citep{larson1969}, increasing by a factor of a few over subsequent years. This is similar in 1-D and 3-D simulations, and with both grey and wavelength-dependent radiative transfer \citep[e.g.][]{bate1998,vaytet2013}. 
\citet{tomida2013aa} reported values an order of magnitude higher: the second core mass was consistently 0.02\solarmass with ideal and non-ideal MHD, and hydrodynamical models. This value might be higher because a portion of the first core had already been accreted onto the second core when the mass was measured. The initial radius of the second core is 1-4\solarrad \citep[e.g.][]{larson1969,masunaga2000,bate2011,bate2014,vaytet2017} but growth is rapid and the second core undergoes repeated periods of transient expansion \citep[e.g.][]{tomida2013aa}.

\citet{bhandare2020} followed the evolution of the second core for over 100 years after its formation for initial protostellar cores with the same range of initial masses as \citet{bhandare2018aa} discussed above ($0.1<M_{\rm{cc}}<100$\solarmass). By the end of the simulations the second core mass was 0.03\solarmass~ for the 0.5\solarmass~ cloud core and $>10$\solarmass~ for initial cores $>80$\solarmass. Massive stars are unlikely to form from the monolithic collapse of a single massive protostellar core \citep[see e.g.][]{tan2014} so the effect of decreasing first core mass for $M_{\rm{cc}}> 10$\solarmass will probably not be observed. However, the mass accretion onto a star--forming core that is being fed by a filament could prevent a stable first core from forming and push the star--forming core straight to second collapse as in \citet{bhandare2018aa}.

Quoting masses for the second core is not particularly informative since the growth is so rapid. What we can say is that it is considerably more compact than the first core and, as such, can launch faster outflows as will be discussed later.

\subsubsection{Discs, pseudo-discs and fragmentation}
\label{sec:morphology}

A rotating core gives rise to an oblate first core with a radius larger than that of a spherical core (10-20~au as opposed to $\sim 5$~au; e.g. \citealt[][]{commercon2011fc,tomida2013aa,bate2014,wurster2021b}). The radius of the first core grows as material is accreted. A rapidly rotating first core can undergo a rotationally-driven bar-mode instability. This gives rise to a gravitationally unstable disc with spiral arms that transport angular momentum outward \citep{bate1998,saigo2008aa,machida2011discs}. \citet{bate2011} refers to these as `pre-stellar' discs because large discs up to $\sim50$~au in radius may be formed before the stellar core is formed. \citet{tsukamoto2015aa} even found a 100~au disc formed in an RHD model. When combined with the infall, the rotational instability may cause the disc to fragment \citep{bate2011}.

The disc is very thick to begin with and becomes thinner as the first core evolves and beyond the formation of the second core. Hydrodynamical simulations showed that the `pre-stellar' disc may have an aspect ratio $H/R \approx$~0.6 – 0.9 which reduces to $H/R<0.1$ after second collapse \citep{machida2010}. When the second core forms, there is more mass contained in the former first core than in the newly formed second core.

Magnetic fields act to reduce rotation \citep{banerjee2006aa,hennebelle2008aa,hennebelle2008b}. Under ideal MHD with a strong magnetic field, the protostellar collapse occurs primarily along the field lines.{\edit The magnetic field lines threaded throughout the protostellar core couple the high density gas near the centre to low density gas further out. Since the gas closer to the centre of the core rotates faster, the field lines become twisted. The twisted field lines generate a torque that slows the rotation of the higher density gas and transfers angular momentum outwards to the lower density gas. This process of magnetic braking prevents the formation of a rotationally-supported disc but a magnetically-supported pseudo-disc develops \citep{matsumoto2004,hennebelle2008aa}}. The pseudo-disc is thicker when the magnetic field is inclined with respect to the rotation axis and a mutual inclination of just 10-20$\degree$ allows a rotationally-supported disc to form even with a strong magnetic field \citep{machida2006_1,hennebelle2009aa}.

Non-ideal MHD processes allow the gas to drift relative to the magnetic field such that discs can form and are able to fragment \citep[e.g.][]{krasnopolsky2011,tomida2013aa,tomida2015,masson2016}. Discs are however prevented from forming when the magnetic field and rotation axis are aligned due to braking from the Hall effect \citep{tsukamoto2015aa,wurster2018c}. \citet{wurster2021b} report the development of a counter-rotating envelope as well as a disc in some models. However, this model implements only the Hall effect, neglecting the other ideal MHD effects and begins with somewhat unrealistic conditions. Counter-rotating envelopes are therefore probably not formed in nature.

In summary, rotationally-supported discs are expected to form in the first and second core phases, and to have similar structures predicted by purely hydrodynamical models albeit with a smaller outer radius.

\subsubsection{Outflows}
\label{sec:outflows}
Outflows have long been proposed as a feature to distinguish sources containing a first core from starless cores. Unfortunately this has produced a few `false alarms' among candidate first cores. In this section I will summarise the predictions for the properties of the first and second core outflows and evaluate the velocities that are most likely.

Simulations predict that the first core outflow is wide and could extend $\sim$~600~au \citep{tomisaka1998,tomisaka2002aa,allen2003aa,banerjee2006aa}, though other models predict an extent of at most a few 10s~au \citep[e.g.][]{machida2005aa,bate2014}. The outflow velocity is consistently predicted to be no more than a few km~s$^{-1}$ and the highest values are $\sim$~5~km~s$^{-1}$ \citep{machida2008aa}. The first core outflow may be slower than the infall velocity ($<1$~km~s$^{-1}$, \citealt{machida2005b}) and comparable to the rotational velocity at the launching radius since it is magneto-centrifugally driven ($\sim$~1~km~s$^{-1}$ at $\sim$~10~au,\citealt{tomida2013aa}). \citet{bate2014} and \citet{wurster2021b} obtain broad, rotating, 1-3~km~s$^{-1}$ outflows, with the rotation velocity similar to the outflow velocity, and infall along the central axis. The extent of the first core outflow depends on the lifetime of the first core \citep{tomida2013aa,tomida2015} and also on the non-ideal MHD processes acting \citep{wurster2021b}. An extent of a few hundred au is very much an upper limit.

In contrast to most other works, the simulations of \citet{price2012ab} produce highly collimated `jets' with velocities of up to 7~km~s$^{-1}$ driven from the first core. However, they note that the ideal MHD method employed is unrealistic for these conditions and therefore these jets are unlikely to form in reality. \citet{lewis2017} also reported similarly fast outflow velocities but point out that they implement a sink particle to represent the first core when the density exceeds $10^{-10}$~g~cm$^{-3}$ and this has an accretion radius of 1~au. Infalling material will therefore have a higher rotation velocity when it reaches the outflow launch radius, boosting the outflow velocity beyond what is realistic \citep{price2003}. With this in mind, the conclusion from the simulations should be that first core outflows have maximum velocities of a few (most likely $<3$) km~s$^{-1}$.

The depth of the second core's potential well is much greater hence it drives a faster outflow than the first core. The radiation hydrodynamical simulations of \citet{bate2010aa,bate2011} showed that the high initial accretion rate onto the newly formed second core leads to rapid heating of the remnants of the first core, launching a slow outflow. This mechanism probably assists the launching of the magnetically-driven outflow. {\edit MHD simulations typically find that the second core drives a two-component outflow. The broader outflow is magneto-centrifugally-driven and contains a fast, narrow outflow, often referred to as a jet, launched by the magnetic pressure from the twisted magnetic field in the disc \citep[e.g.][]{banerjee2006aa,tomisaka2002aa,tomida2010may,machidabasu2019}.} Velocities of the second core outflow obtained in simulations range from 1.5~km~s$^{-1}$ \citep{hennebelle2008aa} to 30-40~km~s$^{-1}$ \citep{machida2006aa,machida2008aa} {\edit and as high as 100~km~s$^{-1}$ within the first 2000 years \citep{machidabasu2019}}. The second core outflow velocity depends on the rotation of the protostellar core: in resistive MHD models, \citet{tomida2013aa} find 6~km~s$^{-1}$ vs 15~km~s$^{-1}$ respectively for slow and fast rotating cores. The outflow velocity is much greater than the maximum rotation velocity at the launch radius, indicating that there is some additional acceleration at work from the magnetic pressure \citep[][]{wurster2021b}.

 It is important to note that outflows do not always form. Firstly, a high accretion rate in the envelope can block the outflow, and this is not always considered in isolated disc models \citep{machida2020,wurster2021b}. Secondly, the outflow is always parallel to the magnetic field \citep{matsumoto2004}, which means that misalignments between the magnetic field and the rotation axis can disrupt the outflow. If the misalignment between rotation and magnetic field is small, a collimated outflow forms but with a 20-45$\degree$ misalignment the magnetic field winds up and disrupts the outflow \citep{lewis2015}. For larger misalignments, the outflow is spherical. Thirdly, a fast outflow is not always produced in non-ideal MHD models \citep{machida2006aa}. The Hall effect suppresses the outflow when the magnetic field is anti-aligned to the rotation axis \citep{wurster2021b}. Finally, we must remember when comparing to observations that we measure the \textit{projected} velocity of the outflow, which will be significantly slower than the maximum outflow speed.

\subsubsection{First core lifetime}

The lifetime of the first core is affected by the strength of the support against gravitational collapse and the rate at which it can cool to keep the temperature below that required to dissociate molecular hydrogen as it is heated by infalling material. Hence rotation, protostellar core mass, magnetic field strength and orientation, and opacity all come into play.

In purely hydrodynamical models, fast rotation leads to a disc-like first core with a lifetime of up to a few thousand years \citep{bate1998,saigo2008aa,bate2011}. In non-ideal MHD models the lifetime is significantly shorter, at $\sim$200-600 years \citep{tomida2013aa,tomida2015,wurster2021b}. \citet{bhandare2018aa} find that the first core lifetime scales as $t_{\rm {fc}}\propto M^{-0.5}$ for protostellar core mass $M_{\rm{cc}}<10$\solarmass~ but steepens to $t_{\rm {fc}}\propto M^{-2.5}$ for $M_{\rm {cc}}>10$\solarmass, due to the additional ram pressure discussed above. Protostellar cores with $M<10$\solarmass~ give rise to first cores with lifetimes of a few hundred years. Beyond this, the first core lifetime is 10-100$\times$ shorter and the resulting first core is therefore essentially unobservable.

First cores that form in very low mass cloud cores (0.1\solarmass) are likely to have far longer lifetimes because they rely on radiative cooling and contraction rather than accretion to reach the central density and temperature required for the second core to form. Most of the envelope is accreted leaving the first core exposed. The increased lifetime, which could be $>10^4$ years, and the reduction in obscuring material means these long-lifetime cores could be as detectable as standard first cores and may even be more luminous at mid-infrared wavelengths \citep{tomida2010dec,stamer2018aa}.
\begin{table}[] 
\begin{threeparttable}
    \centering
     \caption{Summary of predicted properties of the first core. }
    \begin{tabular}{l c c p{2.cm} p{2cm} p{2.cm}}
    \toprule
      & L69$^1$ & 1-D & 3-D rotating hydro & ideal MHD & non-ideal MHD \\
      \midrule
      mass & 0.01 & 0.01-0.04 & 0.005-0.2& 0.02-0.05 & 0.01-0.05\\
      max $R$ (au) & 4  & 7 & 50-100 & 10-20 & 10 \\
      Outflow $v_{\rm{max}}$~(km~s$^{-1}$) & - & - & - & $<7$ & $<3$\\
     Shape & sphere  & sphere & {\raggedright large disc, spiral, fragments} & pseudo-disc, outflow & disc, spiral, fragments, outflow \\
      Max lifetime (years) &  few hundred & few hundred & few 1000 &$\sim 100$ & 200-600\\
      \midrule
    \end{tabular}
  
      \begin{tablenotes}
   \small
   \item $^1$\citet{larson1969}
   \end{tablenotes}
    \label{tab:FHSCproperties}
\end{threeparttable}
\end{table}

\section{Predicted observational characteristics}

In \citet{larson1969}, it was pointed out that the stellar core remains obscured by dust for some time. Since then, simulated observations indicate that there may nevertheless be identifiable features. Estimates of the luminosity of the protostellar core during first and second core stages are a first step to identifying candidate sources. The spectral energy distribution has proved useful for classifying protostars and is another likely diagnostic of early star formation. I first discuss the predictions for the SEDs obtained from numerical simulations before giving an overview of the theoretical work to model molecular line emission and chemical evolution during protostellar collapse.

\subsection{Luminosity and spectral energy distribution}

\begin{figure}
    \centering
    \includegraphics[width=15cm]{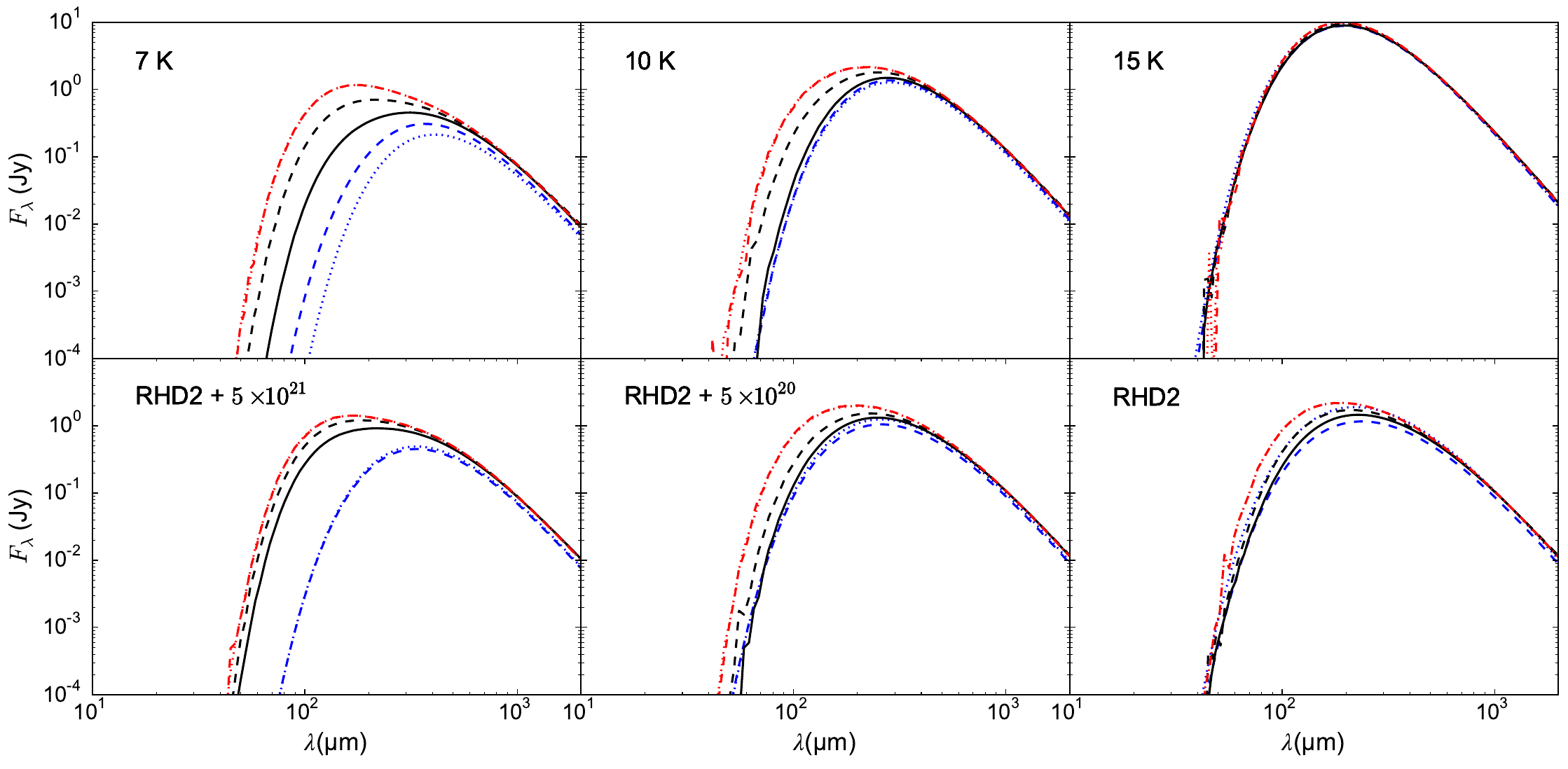}
    \caption{Examples of the simulated spectral energy distribution for key stages during protostellar collapse. The central density was $1.4\times 10^{-18}$\gcc~ (blue dotted), $10^{-12}$\gcc~ (first core formed, blue dashed), $5\times10^{-11}$ (solid black), $10^{-9}$\gcc~ (dashed black), $10{-4}$\gcc~ (second collapse, red dotted), and $10^{-2}$\gcc~ (second core formed, red dashed). The three models in the top panel start from Bonnor-Ebert spheres at the uniform temperatures indicated in the top left corner of each panel. For models in the bottom panel the temperature was initialised from exposure to interstellar radiation, with added attenuation equivalent to an additional column density of \ce{H2} $n_{\rm{H2}}= 5\times10^{21}$ and $5\times10^{20}$ and without attenuation. From \citet{young2018aa}, Fig.~7.}
    \label{fig:SEDfig}
\end{figure}

Models agree that the protostellar core remains very dim throughout the first collapse stage with a luminosity in the range $0.003 \lesssim L_{\rm{fc}} \lesssim 0.1$\solarlum~ \citep[e.g.][]{yorke1993,boss1995,masunaga1998,st2006,commercon2012a,vaytet2013}. However, the earlier estimates of the luminosity that are derived from barotropic models are likely underestimates because they do not consider the significant heating of the former first core region when the second core forms \citep{tomida2010may}. The first core luminosity is therefore more likely to be $L_{\rm{fc}} \gtrsim 0.01$\solarlum.

After second core formation the luminosity increases rapidly, due to the accretion luminosity rather than the intrinsic radiative luminosity of the second core. The emergent luminosity is predicted to be a few tens \solarlum, increasing quickly as the accretion rate increases after second core formation \citep{masunaga2000,jones2018aa}.

The spectral energy distribution (SED) of the first core peaks between 100 and 200 \textmu m \citep[e.g.][]{boss1995,st2006,omukai2007,commercon2012a}. Radiative transfer models of the collapsing protostellar core show there is a gradual shift in the peak of the SED to shorter wavelengths \citep{yorke1993,masunaga2000,saigo2011,commercon2012a,young2018aa}. \citet{commercon2012a} found that there is no clear change after the formation of the first core. Though the change in the SED is gradual, \citet{young2018aa} showed that the first core is distinguishable from a protostellar core undergoing the first collapse phase because the $160$~\textmu m flux rises quickly after first core formation.

We cannot probe the first core directly at wavelengths $\lambda < 850$~\textmu m because the envelope is optically thick and flux emitted by the first core at these wavelengths is completely reprocessed by the time it emerges from the collapsing protostellar core. The shape of the infalling envelope affects the attenuation of the first core flux such that the flattened structure of a rotating core leads to significant emergent flux between 20 and 100~\textmu m when viewed face-on \citep[e.g.][]{bodenheimer1990,boss1995,masunaga1998,commercon2012a,young2018aa}. This is because the optical depth is much lower in the polar direction than perpendicular to the rotation axis. Flux at wavelengths $<100$~\textmu m is therefore highly dependent on the internal structure. The shape of the SED is due to the temperature structure of the warm region outside the first core so the evolution of the SED is due to the heating of the gas and dust surrounding the first core \citep{young2018aa}. The 24~\textmu m flux emerging from the fastest rotating, and therefore most oblate, first cores {\edit was predicted to be} bright enough to be detectable in {\it Spitzer} observations \citep{boss1995,commercon2012a} \footnote{In 1995 {\it Spitzer} was known as SIRTF and subsequently launched in 2003.}{\edit but the optical depth to the observer is probably too high in all but the lowest mass cores \citep{tomida2010dec,young2018aa}. The }70~\textmu m flux could be detectable in {\it Herschel} observations for more evolved, faster rotating cores viewed pole-on \citep{young2018aa}.

There are unlikely to be any changes in the SED of an evolving first core either if it is observed pole-on and it has a dense outflow, or if it is observed edge on \citep{commercon2012a}. Furthermore, there is also no evolution of the SED if the edge of the protostellar core is exposed to the unattenuated interstellar radiation field and therefore heated to $T>10$~K \citep{young2018aa}. This leaves us with something of a catch-22: first cores that are most likely to be detectable because they are less embedded are also unlikely to have a distinctive SED because they are exposed to more heating.

There is no immediate change in the SED after the formation of the second core despite the rapid increase in the central temperature. This is because the second collapse occurs within the former first core so the temperature in the region probed by the SED (the inner envelope) has changed little in this time \citep{young2018aa}.

\subsection{Continuum imaging}

\begin{figure}
    \centering
    \includegraphics[width=15cm]{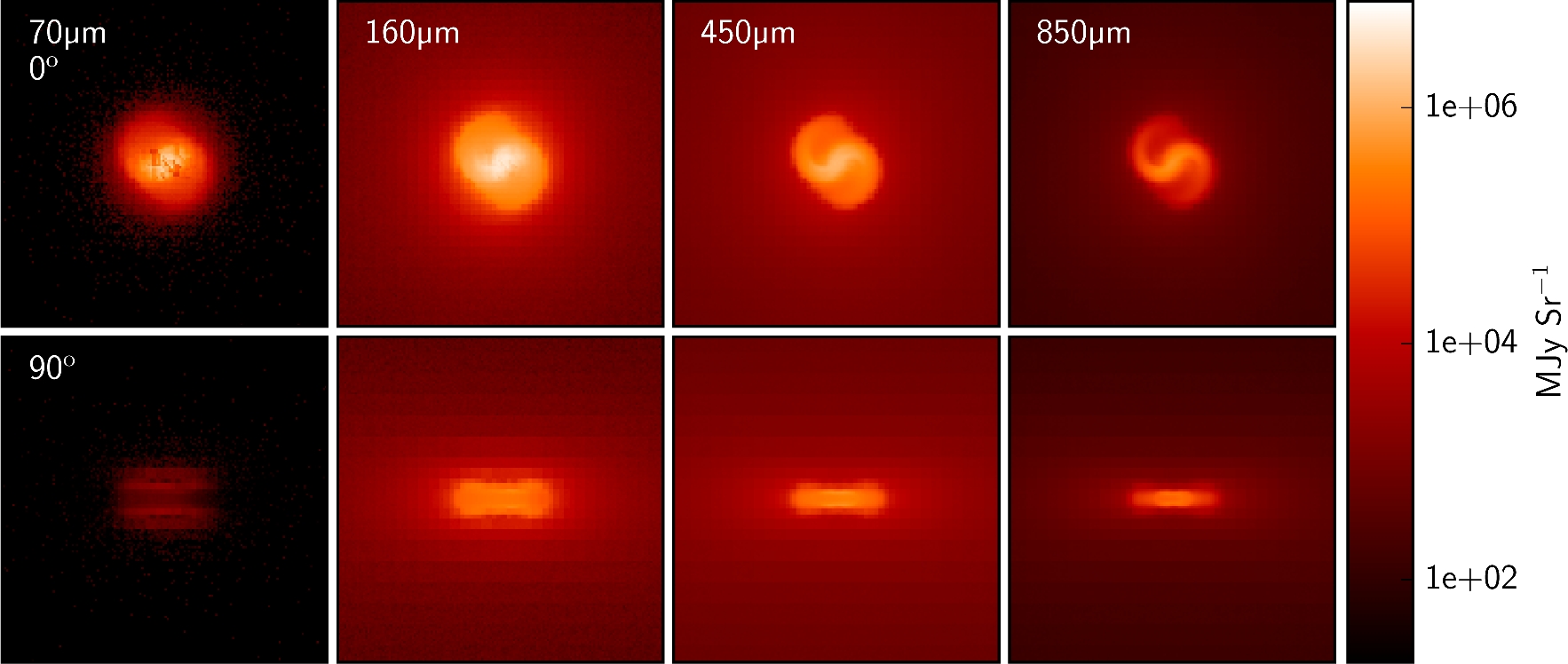}
    \caption{Synthetic observations of a rotating first core developing spiral arms viewed face-on (top row) and edge-on (bottom row), at 4 different wavelengths. \citet{young2018aa} Fig. 6.}
    \label{fig:synthim}
\end{figure}

Since the collapsing envelope is largely optically thick, images of slowly rotating protostellar cores containing a first or second core will be similar to that of starless cores.
However, first and second cores formed in moderately rotating and in magnetized protostellar cores have features that could be detected. The fragments and spiral features associated with fast rotating first and second cores should be detectable at distances of 150~pc with ALMA \citep{saigo2011,commercon2012b}. Synthetic continuum emission maps at different wavelengths are shown in Fig.~\ref{fig:synthim}. In the mid-infrared the structure is unclear. At submillimetre wavelengths (850~\textmu m) the envelope is optically thin and the spiral structure is clear. Simulated emission maps also indicate that the identification of an outflow, pseudo-disc or disc should be possible in ALMA bands 3 and 4, which means that magnetised and non-magnetised cases are distinguishable \citep{commercon2012b}.

\subsection{Molecular line emission and chemistry}

Chemical abundances and gas kinematics are some important properties that can be derived from observations of molecular lines. First and second cores are deeply embedded so optically thin lines are required to probe them and their immediate surroundings. Hydrodynamical models are often used as a starting point for producing synthetic observations of line emission. Sometimes uniform molecular abundance fractions are assumed for post-processing snapshots of the hydrodynamical model with frequency--dependent radiative transfer and sometimes molecular depletion due to e.g. freeze--out is parameterized with a dependence on temperature and/or density. These kinds of models are useful for predicting the line profiles and maps of common molecules.

Measuring the kinematics provides another approach to determining if a source contains a first or second core. Simulated \ce{CS} emission using a uniform \ce{CS} abundance show the signatures of infall, a rotating outflow and an increase in line width compared to during the first collapse \citep{tomisaka2011}. The line profiles of collapsing protostellar cores are sensitive to differences in depletion which means that chemistry should be considered \citep{rawlings2001aa}. Accounting for chemical evolution would lead to stronger differences in the \ce{CS} emission as the core evolves. While discs and pseudo-discs look like identical elongated structures in continuum images, kinematic measurements may tease apart the rotational motions. \ce{CO} isotopologues can reveal the velocity gradients across the observed structure. The velocity gradient is parallel with the disc structure for models of a rotationally supported disc but is perpendicular for models of a pseudo--disc, in which the velocity gradient traces the infalling gas \citep{harsono2015}.

Chemical modelling of a collapsing protostellar core adds an additional stage to modelling observations and a great deal more complexity. The temperature difference between the first core and more evolved sources drives changes in the abundances of many chemical species. Predicting changes in molecular abundances during the collapse of a protostellar core gives an idea of molecules that could be used to distinguish starless cores, first cores, second cores, and more evolved protostars.

\begin{figure}
    \centering
    \includegraphics[width=6cm]{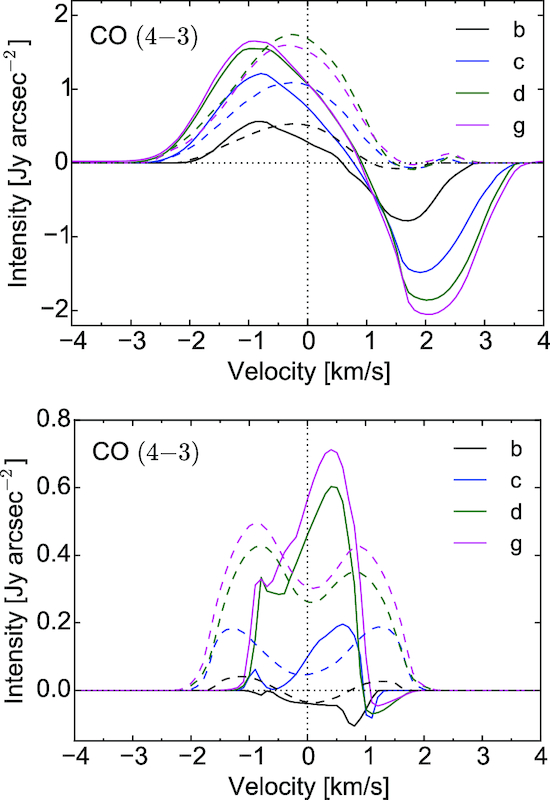}
    \caption{Simulated CO spectra averaged over a central 0.67 arcsec diameter aperture for an evolving first/second core at a distance of 150~pc. The chemical evolution was calculated as a post-process for an RHD model (top) and an ideal MHD model (bottom) of protostellar collapse. Solid lines: viewing inclination $i = 0\degree$ and dashed lines: $i = 90\degree$. Early, mid and late first core stage are designated b,c and d, and g is just after the formation of the second core. From \citet{young2019}.}
    \label{fig:COlineprofile}
\end{figure}

During the isothermal collapse phase, the density increases, which accelerates the freeze-out of most species \citep{aikawa2008aa,van-weeren2009aa,furuya2012aa}. As the centre warms up, CO sublimates and complex organic molecules that formed on the grain surface at low temperatures are released into the gas phase \citep{aikawa2008aa}. The chemical evolution speeds up within the first core and the abundances of many species increase by several orders of magnitude. \ce{CO} and \ce{OCS}, for example, saturate at maximum values very quickly after the formation of the first core. Others such as \ce{CS}, \ce{CN} and \ce{NH_3}, increase in abundance throughout the first core stage \citep{young2019}. \ce{HCN},\ce{HCN}, \ce{H2O}, \ce{HNCO}, \ce{H_2CO}, \ce{HC_3N}, \ce{C_3H_2} and \ce{CH_3OH} are also likely tracers of the first core \citep{hincelin2016}. During the first core stage, the inner envelope surrounding the first core also heats up. In this region, ices frozen onto dust grains are sublimated and gas phase abundances increase. The relative abundance of \ce{HCO^+} decreases in dense cold gas compared to the standard interstellar medium but, after the formation of the first core, the abundance peaks in a shell at the edge of the first core \citep{young2019}. The increase in the abundance of \ce{CO} (and other species) in the inner envelope means the first core may be observed in {\it absorption} against the cooler, optically thick gas \citep{young2019} as is observed in some young protostellar sources \citep[e.g.][]{ohashi2014aa}. The absorption feature is at redshifted velocites $0.5 \lesssim v \lesssim 1.5$~km~s$^{-1}$, because the obscuring gas is infalling onto the first core (see Fig.~\ref{fig:COlineprofile}).

The velocity of the first core outflow is so low that rotation of the outflow may be easier to observe \citep{yamada2009,young2019}. It should be noted that although the first core outflow may extend to a few hundred au, the outflow is cold and probably only the inner few tens of au are warm enough during the first core lifetime for \ce{CO} to be in the gas phase \citep{hincelin2016,young2019}. This means that it is only likely to be possible to detect the first core outflow close to the first core itself. \citet{hincelin2016} suggest that an alternative signature of the first core outflow is the depletion of \ce{CS}, \ce{CN}, \ce{N_2 H^+}, \ce{C_2 H},
and \ce{H_2 CN} relative to the envelope. They additionally note that the molecular abundances of the outflow depend on the magnetic field strength.

After the second core forms the temperature rises quickly in the centre. If an outflow forms, it will sweep up hot gas from the former first core and inner envelope and send this outwards. By around 200 years after second core formation the \ce{CO} sublimation radius shifts outward from $\sim 60$~ au to $\sim 150$~au, perhaps further if the accretion rate is higher \citep{young_thesis}. The rapid increase in the abundances of molecules including \ce{HCN} \citep{lee2004aa}, \ce{CS}, \ce{CN} and \ce{H2CN} in a spatially resolvable region around the second core will make the detectable composition distinct from that of a first core. In particular, \ce{CS} within the protostellar core is probably undetectable during the first core stage but traces the inner disc and envelope region within a couple of hundred years of second core formation \citep{young_thesis}.

Models of chemical evolution are affected by the speed of collapse because of the change in the lifetime of the protostellar core compared to the chemical and freeze-out timescales. Magnetic support increases the lifetime of a protostellar core, allowing more time for freeze-out and therefore changing the chemical evolution \citep{,priestley2018ab,priestley2019}. However, despite extremely large differences in molecular abundances at the centre of protostellar cores with varying levels of magnetic support, the integrated molecular column density is similar, rendering the difference undetectable \citep{priestley2021}. A related finding is that hydrodynamical simulations may overestimate abundances of \ce{HCO+}, \ce{NH3}, \ce{N2H+}, \ce{CN} and \ce{HCN} because of the different collapse timescales \citep{tassis2012jul,yin2021,2022tritsis}.

{\edit
\section{Observations of first cores and the youngest protostars}
\label{sec:obs}

Millimetre interferometers such as \textit{ALMA} probe the youngest embedded objects, unearthing compact sources that are challenging to interpret. These tend to be categorised as either `starless cores' (dense prestellar cores without a compact central object), first core candidates or Class 0 protostars. After second collapse, the second core evolves very quickly into what would be categorised as a Class 0 source so is not an observable object in its own right.

\subsection{Candidate first cores}
 Table~\ref{tab:candidates} lists the 13 candidate first cores and their main characteristics. Although some of these have already been refuted, they are included because there are recent mentions in the literature that still consider them candidates.

\begin{table}[] 
\begin{threeparttable}
    \centering
     \caption{Features of observed candidate first cores from the literature. References are given in the notes below.\\}
    \begin{tabular}{l c  p{4cm}  p{4.3cm} c }
   \toprule
     Source & Mass [\solarmass] & Outflow$^a$ & Features & First core? \\
      \midrule
\\

   Aqu-MM1 &  0.27$^a$ $^{1}$    &   ?    &    SED fitted by fast rotating first core models  $^{2}$   &   ?  \\
 
   B1-bN & 0.36 $^3$ & 4.5~km~s$^{-1}$, few hundred au extent$^3$ & Class 0 protostar & No \\
   
   B1-bS & 0.36 $^3$ & 7~km~s$^{-1}$, few hundred au extent $^{3,4}$ & $\sim$~100 au compact object $^3$ & No\\
   
   CB17~MMS & 0.04 $^{5}$     & (2.5~km~s$^{-1}$ but now attributed to nearby protostar $^6$)  &  Starless core- no compact mm source detected $^6$  &   No \\
   
   Cha-MM1 &  1.44$^{b,7}$    & up to 17~km~s$^{-1}$ $^{8}$,  extends up to 2700~au$^{9}$    &  Probably very young Class 0 protostar   &   No  \\
   HOPS~404 & 0.45 $^{b,10}$ & extent $< 700$~au, $\leq$~2~km~s$^{-1}$  & SED fitted by late first core models $^{11}$ & ? \\
   K242  &   0.4 $^{b,11}$  & ?      &   Fit by FHSC SEDs $^2$       &    ? \\
   L1451-mm & 0.36$^{b,12}$   & 1.5~km~s$^{-1}$, 500~au extent $^{9,13}$   &  Methanol and \ce{SiO} emission detected  $^{21}$   &  No   \\
   L1448 IRS 2E &  (0.04$^{13}$)    & ( 25~km~s$^{-1}$, 9600~au lobe$^{14}$)       & Outflow attributed to different source and there is no star forming core here. $^{9}$   &  No   \\

   MC35-mm & ? & 2-4~km~s$^{-1}$, extends around 2000~au $^{15}$ & \ce{N2D+(3-2)} emission peaks with continuum (cold centre)$^{15}$ & ? \\
   
   Oph A N6 & 0.025–0.03$^{16}$ & Possible slow redshifted lobe$^{16}$ & compact $\sim 100$~au source$^{16}$ & ? \\
   Oph A SM1N & 0.03$^{16}$ & Possible slow, broad, not obviously bipolar$^{16}$ & elongated continuum$^{17}$, unresolved emission $^{16}$ & ? \\
   Per-bolo 45 &   ?   &  No clear outflow$^{9}$    &  Very low luminosity $^{18}$, could be prestellar$^{9}$ &  ?  \\
   Per-bolo 58 &   0.8$^b$  $^{19}$   &  7~km~s$^{-1}$ $^{20}$    & Large central warm region$^{9}$ heated by a central protostar  &   No  \\
  
    \midrule
    \end{tabular}  
     (1)~\citet{Maury:2011aa} (2)~\citet{young2019} (3)~\citet{hirano2014} (4)~\citet{gerin2015}  (5)~\citet{chen2012}  (6)~\citet{spear2021} (7)~\citet{belloche2011a} (8)~\citet{busch2020}  (9)~\citet{maureira2020} 
     (10)~\citet{karnarth2020} (11)~\citet{young_thesis}
     (11)~\citet{Konyves:2015aa}  (12)~\citet{enoch2006} (13)~\citet{pineda2011aa}  (14)~\citet{chen2010} (15)~\citet{fujishiro2020} (16)~\citet{friesen2018aa}  (17)~\citet{friesen2014} (18)~\citet{hatchell2007} (19)~\citet{dunham2011} (20)~\citet{schnee2010} (21)~\citet{wakelam2022}
   \begin{tablenotes}
   \small

   \item $^{a}$  Values here are obtained directly from observations and the source inclinations are not usually known. Hence these correspond to the projection of the actual outflow velocity.
   \item $^{b}$ Including envelope.
   
   \end{tablenotes}
    \label{tab:candidates}
\end{threeparttable}
\end{table}

A couple of former candidates (CB17 MMS and L1448 IRS 2E) were identified due to the presence of outflows and weak continuum detections at 1.3 mm. However, no compact millimetre continuum emission was detected in subsequent deep observations, which means there is no compact object, first core or protostar, at that location and they are `starless cores' \citep{maureira2020,spear2021}. As first cores are fully optically thick objects, with adequate sensitivity and resolution we would always expect to detect a compact source at millimetre wavelengths. The complex density structures of star forming regions can deflect outflows such that their origin is not always obvious. For these candidates, the outflows were later found to be driven by nearby protostars. Conversely, other former candidates were identified due to the lack of outflow associated with a faint continuum source. These, such as Cha-MM1, were subsequently rejected when an extended outflow was detected \citep{busch2020}.

Among the observations of first core candidates, \citet{fujishiro2020} and \citet{maureira2020} demonstrate the application of molecular line observations with ALMA to characterising pre-and protostellar sources. Candidate Per-bolo 58 is rejected because the bright \ce{H^13CO+(1-0)} emission extends around 1000~au across the centre of the source, indicating a large warm region within the core heated by a central protostar \citep{maureira2020}. If the source contained a first core, the warm region where carbon-bearing molecules can be released into the gas phase after being frozen out would be far smaller. In contrast, in Per-bolo 45 the \ce{H^13CO+(1-0)} emission is much fainter in a region within 1000~au of the central source which indicates that the central region is cold enough for carbon-bearing molecules such as CO to freeze out onto dust grains \citep{maureira2020}. Hence, Per-bolo 45 is very unlikely to have evolved beyond the first core stage.

We are now left with eight candidate first cores. Two of these (Aqu-MM1 and K242) require high resolution interferometric observations to search for outflows and to verify that there is a compact object within the source. The remainder have all been observed with ALMA and have a variety of features. We now require observations on sub-100~au scales to probe the density, temperature and composition of very centre of the sources. The outlook for positively identifying the first core in observations is discussed further in section \ref{sec:outlook}.

\subsection{Early evolution: beyond second core formation}
\label{sec:protostarobs}

While observing and firmly identifying a first core remains elusive, progress has been made in characterising very young protostars. Some of these sources are puzzling, having a very low luminosity yet driving an outflow. For example, ten years ago a census of protostars in Orion with {\it Herschel} discovered a number of very young protostars, classified as `PACS Bright Red Sources' or `extreme class 0' objects \citep{stutz2013aa}. Further observations with {\it ALMA} led to the identification of a first core candidate, HOPS~404, among these \citep{karnarth2020}. The other extremely young objects are all deeply embedded young protostars, providing a sample with which to study early outflows. More recently, the dynamics of several very young protostars that do not appear to have evolved far beyond the formation of the second core have been studied. The results include the characterisation of a disc in a system thought to be only around 35,000 years old (HH211, \citealt{lee2023youngdisc}) and some perplexing sources. G208Walma, for example, is a cool, faint object with a significant envelope, similar to how first cores are expected to appear, but it also drives an outflow and a dense jet \citep{dutta2022}. On the other hand, RCrA IRS5N is a Class 0 object that is thought to be too luminous to contain a first core but has no sign of an outflow \citep{sharma2023}.

The chemical structures within pre- and protostellar cores can be explored thanks to {\it ALMA} observations of molecular line emission \citep[e.g.][]{fujishiro2020,maureira2020,caselli2022} and the chemical evolution of star-forming cores is also becoming clearer. Many complex organic molecules have been detected, and `hot corinos', compact warm regions surrounding some of the youngest protostars, are found to be especially rich in these species. This is thought to occur because complex organic molecules form on the surface of dust grains and are released into the gas phase when ices evaporate from the grain surface. Trends in the abundances of complex organic molecules have been found that correlate with the evolutionary stage. For example, the deuteration of methanol increases between the starless core phase and the formation of the protostar, before decreasing again within the hot corino region \citep{aikawa2012}. {\it ALMA} observations of methanol lines illustrate an application by showing that the fraction of deuterated methanol in an `extreme Class 0' source (\citealt{stutz2013aa}; and see section \ref{sec:protostarobs}) appears to mark a transition between that of a prestellar core to an embedded protostar (Fig.~\ref{fig:methfrac}, \citealt{lee2023corino}). Given the youth of the source, the authors propose that the recently-evaporated complex molecular species represent the species formed during the cold phase of the core, that is before the formation of the second core, or protostar. Hence, it may be possible to determine the chemical evolution that occurs during the first core phase through measuring abundances of complex organic molecules in the youngest Class 0 protostars.

\begin{figure}
    \centering
    \includegraphics[width=13cm, trim= 0cm 3.6cm 0cm 0cm, clip]{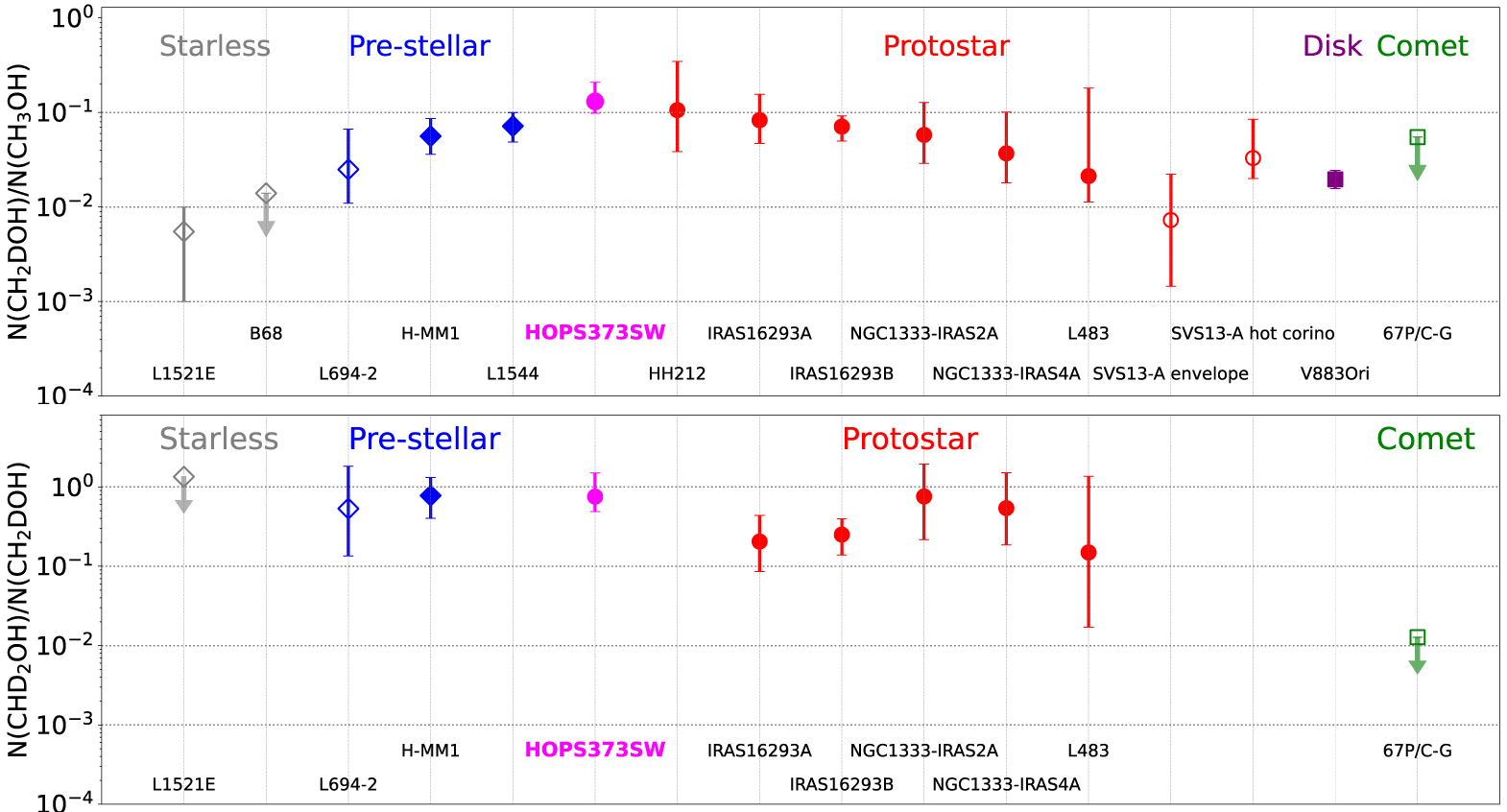}
    \caption{The trend in the fraction of deuterated methanol from starless core to comet as shown by the column density ratio of \ce{CH2DOH}/\ce{CH3OH}. Filled and open symbols indicate single-dish and interferometric observations, respectively. The `extreme Class 0' source HOPS373SW represents a transition between the chemistry of a cold core and of a core in which the central region is warmed by the protostar. Figure adapted from \citet{lee2023corino}, Fig.~9 (CC BY).}
    \label{fig:methfrac}
\end{figure}

It is clearly an exciting time for studying young protostellar systems given the quantity of new data probing their inner regions. These kinds of observations provide useful tests of the dynamical and chemical states of young protostellar systems which will enable us to discriminate between models.}

\section{Discussion}
\label{sec:discussion}
\subsection{What next for modelling?}
\subsubsection{Dust}
\label{sec:dust}
With observations showing significant differences in the dust grain size distribution between the interstellar medium and protoplanetary discs, models are being developed to explore how and when grain growth occurs. We are beginning to explore grain growth during the first core stage and, by extension, the role that this stage plays in planet formation. Even before the first core forms, during the first stage of protostellar collapse, dust grains larger than 100~\textmu m drift inwards and concentrate at the centre such that the dust content near the centre can be enhanced by a factor of $\sim2$ \citep{bate2017aa,lebreuilly2019,lebreuilly2020,koga2022}. Grains may grow by a few tenths of a \textmu m in the infalling envelope and within the first core can grow to bigger than $100$~\textmu m \citep{bate2022,lebreuilly2023}. Dust grains are likely to be swept up in the outflow and then may fall back down on to the disc at larger radii \citep{koga2022}. These results paint the picture that the first core probably plays an important role in growing dust grains that will populate the protoplanetary disc. \citet{bate2022} suggests too that the formation of grains in the first core could explain the origin of calcium rich inclusions. These studies are just the beginning. Models of grain growth during the protostellar collapse have not yet followed the evolution beyond the first core. Dust grains sublimate at $T\sim 1500$~K, which is a few hundred K lower than the hydrogen dissociation temperature that triggers the second collapse. The next step would be to extend these models to assess the survival of the large grains produced in the first core and how far into the disc and outflow they are distributed.

\subsubsection{Chemistry}
{\edit 

Measurements of molecular abundances and the distribution of molecular species within protostellar cores are providing a new insight into the evolutionary stage of young sources (see section \ref{sec:obs}). The detection of molecular lines that trace warm cores has been used to discount first core candidates and so it is important that models of first core chemistry, and the simulated line maps, are sound. The finding that molecular line emission of young hot corinos may indirectly reveal the chemistry associated with the earlier, first core stage has useful implications. This strongly motivates new chemical models of the first and second core stages that include complex organic molecules, now that we have an observational test. This may reveal whether the first core and/or the formation of the second core play a role in enriching protostellar cores with complex organic molecules.

Chemical evolution is generally only modelled for isolated protostellar cores but there are predicted to be differences in the chemistry between embedded and isolated prestellar cores \citep{priestley2023}. This needs to be considered because these differences could propagate through to change the later chemical evolution within the protostellar core and also because the chemistry in the outer envelope evolves very slowly along with the physical properties. Observations that probe the envelope with therefore be sensitive to the initial differences in abundance that are due to the environment. Synthetic observations of key molecular lines for evolving protostellar cores in both isolated and embedded environments would allow a more robust interpretation of the ages of these sources.

The significant grain growth and increase in dust to gas ratio in the dense regions described in section \ref{sec:dust} also has implications for the chemistry. A higher abundance of dust grains will increase freezeout but could also enhance the formation of complex organic molecules, which are thought to form on the icy surfaces of dust grains. Current models of the chemical evolution during protostellar collapse do not include grain sizes $\gtrsim100$~\textmu m. Chemical models incorporating the changing dust abundance and grain sizes in a collapsing protostellar are required to fully understand the gas composition. Furthermore, the initial composition of the protoplanetary disc is set by the first core and inner envelope so a better understanding of their chemistry would greatly benefit disc models.}

\subsection{Outlook for comparison to observations}
\label{sec:outlook}

Larson realised that the excess infra-red emission detected from T Tauri stars was consistent with them being young stars, still accompanied by infalling dust. He also compared infall velocities with the redshifted absorption lines and found them consistent. Unresolved observations cannot usually distinguish between features like discs and envelopes and give scant information as to the inclination and morphology of a source. \citet{boss1995} asserted that a first core should be detectable at a distance of 100~pc with ISO (launched in 1995) and {\it Spitzer}. Furthermore, with the advent of ALMA, detection of the first core seemed within reach. Ten years on, many former `candidate first cores' have been eliminated and few new potential sources have been identified (See table~\ref{tab:candidates}). {\edit {\it Spitzer} observations have however proved useful in excluding the more evolved sources as it has become clearer that first cores will be very faint at wavelengths $<70$~\textmu m.}

A reason why there were previously more candidate first cores was because of the generous interpretation of outflow velocities predicted by simulations. A low luminosity source is typically designated a `candidate first core' if its luminosity is $L<0.1$\solarlum, it is faint or undetected at $\lambda<70$~\textmu m and no outflow faster than 10~km~s$^{-1}$ is observed. Given the results described in section \ref{sec:outflows}, the latter constraint must be tightened significantly and it seems sensible to exclude sources for which clear outflows are observed. Simulated observations show that there should still be other characteristics that can distinguish first cores from starless cores. For example, chemical evolution may be evident and the visibility amplitude of interferometric observations also shows differences \citep{tomida2010dec}.

There have now been several deep surveys of young stellar objects in all of the nearby star forming regions so why hasn't a first core been detected yet? Estimates of the expected population of first cores have been made by comparing their lifetimes to the estimated lifetimes of Class 0 and Class I objects. The lifetime of Class 0 protostars is estimated to be somewhere between a few $\times 10^4$~years \citep{andre2000,Maury:2011aa} and $1-3\times 10^5$~years \citep{evans2009,Dunham:2015aa}.
With lifetimes of a few hundred years, we can expect the fraction of first cores compared to Class 0 objects to be 0.002-0.02. The young stellar object catalogue of \citet{Dunham:2015aa} includes around 103 Class 0 protostars. Based on this we would expect to find at most 1 or 2 first cores in total in the nearby star forming regions. If the Class 0 lifetime is at the upper end of the range of estimates, we would be unlikely to find any. {\edit (See \citet{maureira2020} for similar discussion.)}

The now-cancelled Space Infrared Telescope for Cosmology and Astrophysics (SPICA) would have had the sensitivity to detect the first core in far-infrared continuum and H$_2$O emission out to 1~kpc \citep{omukai2007}. The MIRI instrument onboard the \textit{James Webb Space Telescope} operates at 5-28~\textmu m and could potentially detect flux from face-on oblate first cores, but a dense outflow may however obscure the flux at such short wavelengths. Though it may not be practically possible to observe objects in the first core stage due to the short lifetimes and lack of current far-infrared instruments, we may be able to infer its existence and properties nonetheless. The role that the first core appears to have in growing dust grains and forming a wide range of molecules may mean that we can make some estimates of its properties by comparing the dust grain and molecule distributions in Class 0 discs and outflows with model predictions. It could be worthwhile modelling the chemistry and dust evolution beyond second core formation to explore the variation in dust and molecular abundances and distribution with the properties of the protostellar core.

{\edit Nevertheless, we are left with 5 candidate first cores that have been observed with {\it ALMA}. What observations do we need to positively identify any of these as a first core? A low luminosity alone is insufficient because deeply embedded protostars and protostars with edge-on discs have also been found to be exceptionally faint. A slow outflow can be indicative, however this is not a necessary criterion which means that Per-bolo~45 should not be discounted. If the magnetic field and rotation axis are misaligned the first core, or indeed protostar, may not drive a typical outflow. Also, as we saw in the case of CB17 MMS, the uneven density structure of star forming regions can stem or redirect outflows in an unexpected direction. The detection of a source in millimetre continuum that shows no compact source smaller than a few au from analysis of the visibility amplitudes would strongly support identification as a first core. Signs of chemical evolution in a dense core that indicate a compact warm central region that is also too cool to be heated by a protostar are another strong indicator.}

{\edit What, then, is required from theory to find the first core? In the {\it ALMA} era, simulated observations should now focus on visibility amplitudes and molecular lines at high angular resolution. Chemistry has become another way of assessing the evolutionary state of a source and the molecular structure probably changes very quickly following second core formation \citep{young_thesis}. Consequently, more work is needed to clarify the chemical changes that occur during and after the formation of the second core and to predict the corresponding observables. We need to clarify the expected observational characteristics of first core outflows. In particular, simulations need to determine the observable extent of first core outflows: if the outflow does extend $<100$~au within the first core lifetime do molecular gas tracers like CO remain in the gas phase throughout the whole outflow? If so, L1451-mm and MC35-mm remain strong candidates.}

\subsection{Conclusion and outlook}

The evolution of a collapsing protostellar core has been studied thoroughly over the last few decades. Given how little the theory has qualitatively changed with the consideration of additional physics in simulations, we can be confident of the central predictions of first and second core properties. A caveat is perhaps that our understanding is based on numerical models that all follow the collapse of isolated protostellar cores. Observations show that cores form in filaments and there may be ongoing accretion onto collapsing protostellar cores which could affect, for example, the first core lifetime.

An important development has been to show that discs can form early, during the first core stage, even in MHD simulations once nonideal process are taken into account. For first core outflows, the consensus from MHD simulations and chemical models is that they are probably very difficult to detect. However, these outflows could play a significant role in redistributing dust grains and molecular species from the first core to the envelope and disc. It could be, however, that the material forming the disc when the second core forms is all accreted very quickly. Further modelling of the evolution of the protostellar core beyond second core formation is needed to follow the development of the disc.

Now that protostellar collapse is well understood we can turn to studying how this phase shapes the resulting protostar system. We are beginning to see that the dust grain size distribution evolves significantly during protostellar collapse, particularly within the first core, and outflows redistribute large grains out into the disc and envelope. This has a number of implications that warrant further study. How does the changing dust composition affect the thermal properties of the first core, for example? Do the large dust grains survive the second core formation and do they seed grain growth in the young protostellar disc? Do we expect variations in the protostellar collapse process to lead to differences in the dust grain distribution in protostellar discs? Similarly, the warm first core seems to be an effective chemical factory. Further open questions are therefore: do these molecules survive beyond second core formation and how does this influence the composition of Class 0 discs and envelopes? 

{\edit
Finally, a positive observational identification of the first core remains difficult but is potentially within reach with the exceptional sensitivity and resolving power of {\it ALMA}. I have listed the remaining candidate first cores and there remains some tension between the predicted extent of the first core outflow and the outflow properties of first core candidates. We now require more detailed predictions of the chemical evolution of the envelope and outflow in order to firmly identify the first core for the first time.}

\section*{Conflict of Interest Statement}

The authors declare that the research was conducted in the absence of any commercial or financial relationships that could be construed as a potential conflict of interest.

\section*{Author Contributions}
AY: Writing–original draft, Writing–review and editing.

\section*{Funding}
AY is grateful for support from the UK STFC via grant ST/V000594/1. 

\section*{Acknowledgments}
I thank Matthew Bate, Ken Rice and the reviewers for helpful comments on the manuscript. I am very grateful to Mar\'{i}a Jos\'{e} Maureira for suggestions and comments regarding observations of first core candidates.


\bibliographystyle{Frontiers-Harvard} 
\bibliography{papers}

\begin{thebibliography}{129}
\providecommand{\natexlab}[1]{#1}
\expandafter\ifx\csname urlstyle\endcsname\relax
  \providecommand{\doi}[1]{doi:\discretionary{}{}{}#1}\else
  \providecommand{\doi}{doi:\discretionary{}{}{}\begingroup
  \urlstyle{rm}\Url}\fi
\providecommand{\selectlanguage}[1]{\relax}
\providecommand{\bibAnnoteFile}[1]{%
  \IfFileExists{#1}{\begin{quotation}\noindent\textsc{Key:} #1\\
  \textsc{Annotation:}\ \input{#1}\end{quotation}}{}}
\providecommand{\bibAnnote}[2]{%
  \begin{quotation}\noindent\textsc{Key:} #1\\
  \textsc{Annotation:}\ #2\end{quotation}}

\bibitem[{{Aikawa} et~al.(2008){Aikawa}, {Wakelam}, {Garrod}, and
  {Herbst}}]{aikawa2008aa}
{Aikawa}, Y., {Wakelam}, V., {Garrod}, R.~T., and {Herbst}, E. (2008).
\newblock {Molecular Evolution and Star Formation: From Prestellar Cores to
  Protostellar Cores}.
\newblock \emph{\apj} 674, 984-996.
\newblock \doi{10.1086/524096}
\bibAnnoteFile{aikawa2008aa}

\bibitem[{{Aikawa} et~al.(2012){Aikawa}, {Wakelam}, {Hersant}, {Garrod}, and
  {Herbst}}]{aikawa2012}
{Aikawa}, Y., {Wakelam}, V., {Hersant}, F., {Garrod}, R.~T., and {Herbst}, E.
  (2012).
\newblock {From Prestellar to Protostellar Cores. II. Time Dependence and
  Deuterium Fractionation}.
\newblock \emph{\apj} 760, 40.
\newblock \doi{10.1088/0004-637X/760/1/40}
\bibAnnoteFile{aikawa2012}

\bibitem[{{Allen} et~al.(2003){Allen}, {Li}, and {Shu}}]{allen2003aa}
{Allen}, A., {Li}, Z.-Y., and {Shu}, F.~H. (2003).
\newblock {Collapse of Magnetized Singular Isothermal Toroids. II. Rotation and
  Magnetic Braking}.
\newblock \emph{\apj} 599, 363--379.
\newblock \doi{10.1086/379243}
\bibAnnoteFile{allen2003aa}

\bibitem[{{Andre} et~al.(2000){Andre}, {Ward-Thompson}, and
  {Barsony}}]{andre2000}
{Andre}, P., {Ward-Thompson}, D., and {Barsony}, M. (2000).
\newblock {From Prestellar Cores to Protostars: the Initial Conditions of Star
  Formation}.
\newblock In \emph{Protostars and Planets IV}, eds. V.~{Mannings}, A.~P.
  {Boss}, and S.~S. {Russell}. 59.
\newblock \doi{10.48550/arXiv.astro-ph/9903284}
\bibAnnoteFile{andre2000}

\bibitem[{{Banerjee} and {Pudritz}(2006)}]{banerjee2006aa}
{Banerjee}, R. and {Pudritz}, R.~E. (2006).
\newblock {Outflows and Jets from Collapsing Magnetized Cloud Cores}.
\newblock \emph{\apj} 641, 949--960.
\newblock \doi{10.1086/500496}
\bibAnnoteFile{banerjee2006aa}

\bibitem[{{Bate}(1998)}]{bate1998}
{Bate}, M.~R. (1998).
\newblock {Collapse of a Molecular Cloud Core to Stellar Densities: The First
  Three-dimensional Calculations}.
\newblock \emph{\apjl} 508, L95--L98.
\newblock \doi{10.1086/311719}
\bibAnnoteFile{bate1998}

\bibitem[{{Bate}(2010)}]{bate2010aa}
{Bate}, M.~R. (2010).
\newblock {Collapse of a molecular cloud core to stellar densities: the
  radiative impact of stellar core formation on the circumstellar disc}.
\newblock \emph{\mnras} 404, L79--L83.
\newblock \doi{10.1111/j.1745-3933.2010.00839.x}
\bibAnnoteFile{bate2010aa}

\bibitem[{{Bate}(2011)}]{bate2011}
{Bate}, M.~R. (2011).
\newblock {Collapse of a molecular cloud core to stellar densities: the
  formation and evolution of pre-stellar discs}.
\newblock \emph{\mnras} 417, 2036--2056.
\newblock \doi{10.1111/j.1365-2966.2011.19386.x}
\bibAnnoteFile{bate2011}

\bibitem[{{Bate}(2022)}]{bate2022}
{Bate}, M.~R. (2022).
\newblock {Dust coagulation during the early stages of star formation:
  molecular cloud collapse and first hydrostatic core evolution}.
\newblock \emph{\mnras} 514, 2145--2161.
\newblock \doi{10.1093/mnras/stac1391}
\bibAnnoteFile{bate2022}

\bibitem[{{Bate} and {Lor{\'e}n-Aguilar}(2017)}]{bate2017aa}
{Bate}, M.~R. and {Lor{\'e}n-Aguilar}, P. (2017).
\newblock {On the dynamics of dust during protostellar collapse}.
\newblock \emph{\mnras} 465, 1089--1094.
\newblock \doi{10.1093/mnras/stw2853}
\bibAnnoteFile{bate2017aa}

\bibitem[{{Bate} et~al.(2014){Bate}, {Tricco}, and {Price}}]{bate2014}
{Bate}, M.~R., {Tricco}, T.~S., and {Price}, D.~J. (2014).
\newblock {Collapse of a molecular cloud core to stellar densities:
  stellar-core and outflow formation in radiation magnetohydrodynamic
  simulations}.
\newblock \emph{\mnras} 437, 77--95.
\newblock \doi{10.1093/mnras/stt1865}
\bibAnnoteFile{bate2014}

\bibitem[{{Belloche} et~al.(2011){Belloche}, {Schuller}, {Parise}, {Andr{\'e}},
  {Hatchell}, {J{\o}rgensen} et~al.}]{belloche2011a}
{Belloche}, A., {Schuller}, F., {Parise}, B., {Andr{\'e}}, P., {Hatchell}, J.,
  {J{\o}rgensen}, J.~K., et~al. (2011).
\newblock {The end of star formation in Chamaeleon I?. A LABOCA census of
  starless and protostellar cores}.
\newblock \emph{\aap} 527, A145.
\newblock \doi{10.1051/0004-6361/201015733}
\bibAnnoteFile{belloche2011a}

\bibitem[{{Bhandare} et~al.(2020){Bhandare}, {Kuiper}, {Henning}, {Fendt},
  {Flock}, and {Marleau}}]{bhandare2020}
{Bhandare}, A., {Kuiper}, R., {Henning}, T., {Fendt}, C., {Flock}, M., and
  {Marleau}, G.-D. (2020).
\newblock {Birth of convective low-mass to high-mass second Larson cores}.
\newblock \emph{\aap} 638, A86.
\newblock \doi{10.1051/0004-6361/201937029}
\bibAnnoteFile{bhandare2020}

\bibitem[{{Bhandare} et~al.(2018){Bhandare}, {Kuiper}, {Henning}, {Fendt},
  {Marleau}, and {K{\"o}lligan}}]{bhandare2018aa}
{Bhandare}, A., {Kuiper}, R., {Henning}, T., {Fendt}, C., {Marleau}, G.-D., and
  {K{\"o}lligan}, A. (2018).
\newblock {First Core Properties: From Low- to High-mass Star Formation}.
\newblock \emph{\aap} 618, A95.
\newblock \doi{10.1051/0004-6361/201832635}
\bibAnnoteFile{bhandare2018aa}

\bibitem[{{Bodenheimer} et~al.(1990){Bodenheimer}, {Yorke}, {Rozyczka}, and
  {Tohline}}]{bodenheimer1990}
{Bodenheimer}, P., {Yorke}, H.~W., {Rozyczka}, M., and {Tohline}, J.~E. (1990).
\newblock {The Formation Phase of the Solar Nebula}.
\newblock \emph{\apj} 355, 651.
\newblock \doi{10.1086/168798}
\bibAnnoteFile{bodenheimer1990}

\bibitem[{{Boss}(1989)}]{boss1989nov}
{Boss}, A.~P. (1989).
\newblock {Protostellar Formation in Rotating Interstellar Clouds. VIII. Inner
  Core Formation}.
\newblock \emph{\apj} 346, 336.
\newblock \doi{10.1086/168014}
\bibAnnoteFile{boss1989nov}

\bibitem[{{Boss} and {Yorke}(1995)}]{boss1995}
{Boss}, A.~P. and {Yorke}, H.~W. (1995).
\newblock {Spectral energy of first protostellar cores: Detecting 'class -I'
  protostars with ISO and SIRTF}.
\newblock \emph{\apjl} 439, L55--L58.
\newblock \doi{10.1086/187743}
\bibAnnoteFile{boss1995}

\bibitem[{{Busch} et~al.(2020){Busch}, {Belloche}, {Cabrit}, {Hennebelle}, and
  {Commer{\c{c}}on}}]{busch2020}
{Busch}, L.~A., {Belloche}, A., {Cabrit}, S., {Hennebelle}, P., and
  {Commer{\c{c}}on}, B. (2020).
\newblock {The dynamically young outflow of the Class 0 protostar Cha-MMS1}.
\newblock \emph{\aap} 633, A126.
\newblock \doi{10.1051/0004-6361/201936432}
\bibAnnoteFile{busch2020}

\bibitem[{{Caselli} et~al.(2022){Caselli}, {Pineda}, {Sipil{\"a}}, {Zhao},
  {Redaelli}, {Spezzano} et~al.}]{caselli2022}
{Caselli}, P., {Pineda}, J.~E., {Sipil{\"a}}, O., {Zhao}, B., {Redaelli}, E.,
  {Spezzano}, S., et~al. (2022).
\newblock {The Central 1000 au of a Prestellar Core Revealed with ALMA. II.
  Almost Complete Freeze-out}.
\newblock \emph{\apj} 929, 13.
\newblock \doi{10.3847/1538-4357/ac5913}
\bibAnnoteFile{caselli2022}

\bibitem[{{Chen} et~al.(2012){Chen}, {Arce}, {Dunham}, {Zhang}, {Bourke},
  {Launhardt} et~al.}]{chen2012}
{Chen}, X., {Arce}, H.~G., {Dunham}, M.~M., {Zhang}, Q., {Bourke}, T.~L.,
  {Launhardt}, R., et~al. (2012).
\newblock {Submillimeter Array and Spitzer Observations of Bok Globule CB 17: A
  Candidate First Hydrostatic Core?}
\newblock \emph{\apj} 751, 89.
\newblock \doi{10.1088/0004-637X/751/2/89}
\bibAnnoteFile{chen2012}

\bibitem[{{Chen} et~al.(2010){Chen}, {Arce}, {Zhang}, {Bourke}, {Launhardt},
  {Schmalzl} et~al.}]{chen2010}
{Chen}, X., {Arce}, H.~G., {Zhang}, Q., {Bourke}, T.~L., {Launhardt}, R.,
  {Schmalzl}, M., et~al. (2010).
\newblock {L1448 IRS2E: A Candidate First Hydrostatic Core}.
\newblock \emph{\apj} 715, 1344--1351.
\newblock \doi{10.1088/0004-637X/715/2/1344}
\bibAnnoteFile{chen2010}

\bibitem[{{Commer{\c{c}}on} et~al.(2011){Commer{\c{c}}on}, {Audit}, {Chabrier},
  and {Chi{\`e}ze}}]{commercon2011fc}
{Commer{\c{c}}on}, B., {Audit}, E., {Chabrier}, G., and {Chi{\`e}ze}, J.~P.
  (2011).
\newblock {Physical and radiative properties of the first-core accretion
  shock}.
\newblock \emph{\aap} 530, A13.
\newblock \doi{10.1051/0004-6361/201016213}
\bibAnnoteFile{commercon2011fc}

\bibitem[{{Commer{\c{c}}on} et~al.(2010){Commer{\c{c}}on}, {Hennebelle},
  {Audit}, {Chabrier}, and {Teyssier}}]{commercon2010aa}
{Commer{\c{c}}on}, B., {Hennebelle}, P., {Audit}, E., {Chabrier}, G., and
  {Teyssier}, R. (2010).
\newblock {Protostellar collapse: radiative and magnetic feedbacks on
  small-scale fragmentation}.
\newblock \emph{\aap} 510, L3.
\newblock \doi{10.1051/0004-6361/200913597}
\bibAnnoteFile{commercon2010aa}

\bibitem[{{Commer{\c{c}}on} et~al.(2012){Commer{\c{c}}on}, {Launhardt},
  {Dullemond}, and {Henning}}]{commercon2012a}
{Commer{\c{c}}on}, B., {Launhardt}, R., {Dullemond}, C., and {Henning}, T.
  (2012).
\newblock {Synthetic observations of first hydrostatic cores in collapsing
  low-mass dense cores. I. Spectral energy distributions and evolutionary
  sequence}.
\newblock \emph{\aap} 545, A98.
\newblock \doi{10.1051/0004-6361/201118706}
\bibAnnoteFile{commercon2012a}

\bibitem[{{Commer\c{c}on} et~al.(2012){Commer\c{c}on}, {Levrier}, {Maury},
  {Henning}, and {Launhardt}}]{commercon2012b}
{Commer\c{c}on}, B., {Levrier}, F., {Maury}, A.~J., {Henning}, T., and
  {Launhardt}, R. (2012).
\newblock Synthetic observations of first hydrostatic cores in collapsing
  low-mass dense cores. ii. simulated alma dust emission maps.
\newblock \emph{\aap} 548, A39.
\newblock \doi{10.1051/0004-6361/201220067}
\bibAnnoteFile{commercon2012b}

\bibitem[{{di Francesco} et~al.(2007){di Francesco}, {Evans}, {Caselli},
  {Myers}, {Shirley}, {Aikawa} et~al.}]{ppVdensecores}
{di Francesco}, J., {Evans}, N.~J., II, {Caselli}, P., {Myers}, P.~C.,
  {Shirley}, Y., {Aikawa}, Y., et~al. (2007).
\newblock \emph{Protostars and Planets V} (University of Arizona Press), chap.
  An Observational Perspective of Low-Mass Dense Cores I: Internal Physical and
  Chemical Properties.
\newblock 17--32
\bibAnnoteFile{ppVdensecores}

\bibitem[{{Dunham} et~al.(2015){Dunham}, {Allen}, {Evans}, {Broekhoven-Fiene},
  {Cieza}, {Di Francesco} et~al.}]{Dunham:2015aa}
{Dunham}, M.~M., {Allen}, L.~E., {Evans}, N.~J., II, {Broekhoven-Fiene}, H.,
  {Cieza}, L.~A., {Di Francesco}, J., et~al. (2015).
\newblock {Young Stellar Objects in the Gould Belt}.
\newblock \emph{\apjs} 220, 11.
\newblock \doi{10.1088/0067-0049/220/1/11}
\bibAnnoteFile{Dunham:2015aa}

\bibitem[{{Dunham} et~al.(2011){Dunham}, {Chen}, {Arce}, {Bourke}, {Schnee},
  and {Enoch}}]{dunham2011}
{Dunham}, M.~M., {Chen}, X., {Arce}, H.~G., {Bourke}, T.~L., {Schnee}, S., and
  {Enoch}, M.~L. (2011).
\newblock {Detection of a Bipolar Molecular Outflow Driven by a Candidate First
  Hydrostatic Core}.
\newblock \emph{\apj} 742, 1.
\newblock \doi{10.1088/0004-637X/742/1/1}
\bibAnnoteFile{dunham2011}

\bibitem[{{Dutta} et~al.(2022){Dutta}, {Lee}, {Hirano}, {Liu}, {Johnstone},
  {Liu} et~al.}]{dutta2022}
{Dutta}, S., {Lee}, C.-F., {Hirano}, N., {Liu}, T., {Johnstone}, D., {Liu},
  S.-Y., et~al. (2022).
\newblock {ALMA Survey of Orion Planck Galactic Cold Clumps (ALMASOP): Evidence
  for a Molecular Jet Launched at an Unprecedented Early Phase of Protostellar
  Evolution}.
\newblock \emph{\apj} 931, 130.
\newblock \doi{10.3847/1538-4357/ac67a1}
\bibAnnoteFile{dutta2022}

\bibitem[{{Enoch} et~al.(2006){Enoch}, {Young}, {Glenn}, {Evans}, {Golwala},
  {Sargent} et~al.}]{enoch2006}
{Enoch}, M.~L., {Young}, K.~E., {Glenn}, J., {Evans}, N.~J., II, {Golwala}, S.,
  {Sargent}, A.~I., et~al. (2006).
\newblock {Bolocam Survey for 1.1 mm Dust Continuum Emission in the c2d Legacy
  Clouds. I. Perseus}.
\newblock \emph{\apj} 638, 293--313.
\newblock \doi{10.1086/498678}
\bibAnnoteFile{enoch2006}

\bibitem[{{Evans} et~al.(2009){Evans}, {Dunham}, {J{\o}rgensen}, {Enoch},
  {Mer{\'\i}n}, {van Dishoeck} et~al.}]{evans2009}
{Evans}, I., Neal~J., {Dunham}, M.~M., {J{\o}rgensen}, J.~K., {Enoch}, M.~L.,
  {Mer{\'\i}n}, B., {van Dishoeck}, E.~F., et~al. (2009).
\newblock {The Spitzer c2d Legacy Results: Star-Formation Rates and
  Efficiencies; Evolution and Lifetimes}.
\newblock \emph{\apjs} 181, 321--350.
\newblock \doi{10.1088/0067-0049/181/2/321}
\bibAnnoteFile{evans2009}

\bibitem[{{Friesen} et~al.(2014){Friesen}, {Di Francesco}, {Bourke}, {Caselli},
  {J{\o}rgensen}, {Pineda} et~al.}]{friesen2014}
{Friesen}, R.~K., {Di Francesco}, J., {Bourke}, T.~L., {Caselli}, P.,
  {J{\o}rgensen}, J.~K., {Pineda}, J.~E., et~al. (2014).
\newblock {Revealing H$_{2}$D$^{+}$ Depletion and Compact Structure in Starless
  and Protostellar Cores with ALMA}.
\newblock \emph{\apj} 797, 27.
\newblock \doi{10.1088/0004-637X/797/1/27}
\bibAnnoteFile{friesen2014}

\bibitem[{{Friesen} et~al.(2018){Friesen}, {Pon}, {Bourke}, {Caselli}, {Di
  Francesco}, {J{\o}rgensen} et~al.}]{friesen2018aa}
{Friesen}, R.~K., {Pon}, A., {Bourke}, T.~L., {Caselli}, P., {Di Francesco},
  J., {J{\o}rgensen}, J.~K., et~al. (2018).
\newblock {ALMA detections of the youngest protostars in Ophiuchus}.
\newblock \emph{\apj} 869, 158.
\newblock \doi{10.3847/1538-4357/aaeff5}
\bibAnnoteFile{friesen2018aa}

\bibitem[{{Fujishiro} et~al.(2020){Fujishiro}, {Tokuda}, {Tachihara},
  {Takashima}, {Fukui}, {Zahorecz} et~al.}]{fujishiro2020}
{Fujishiro}, K., {Tokuda}, K., {Tachihara}, K., {Takashima}, T., {Fukui}, Y.,
  {Zahorecz}, S., et~al. (2020).
\newblock {A Low-velocity Bipolar Outflow from a Deeply Embedded Object in
  Taurus Revealed by the Atacama Compact Array}.
\newblock \emph{\apjl} 899, L10.
\newblock \doi{10.3847/2041-8213/ab9ca8}
\bibAnnoteFile{fujishiro2020}

\bibitem[{{Furuya} et~al.(2012){Furuya}, {Aikawa}, {Tomida}, {Matsumoto},
  {Saigo}, {Tomisaka} et~al.}]{furuya2012aa}
{Furuya}, K., {Aikawa}, Y., {Tomida}, K., {Matsumoto}, T., {Saigo}, K.,
  {Tomisaka}, K., et~al. (2012).
\newblock {Chemistry in the First Hydrostatic Core Stage by Adopting
  Three-dimensional Radiation Hydrodynamic Simulations}.
\newblock \emph{\apj} 758, 86.
\newblock \doi{10.1088/0004-637X/758/2/86}
\bibAnnoteFile{furuya2012aa}

\bibitem[{{Galli} and {Shu}(1993)}]{galli1993}
{Galli}, D. and {Shu}, F.~H. (1993).
\newblock {Collapse of Magnetized Molecular Cloud Cores. I. Semianalytical
  Solution}.
\newblock \emph{\apj} 417, 220.
\newblock \doi{10.1086/173305}
\bibAnnoteFile{galli1993}

\bibitem[{{Gerin} et~al.(2015){Gerin}, {Pety}, {Fuente}, {Cernicharo},
  {Commer{\c c}on}, and {Marcelino}}]{gerin2015}
{Gerin}, M., {Pety}, J., {Fuente}, A., {Cernicharo}, J., {Commer{\c c}on}, B.,
  and {Marcelino}, N. (2015).
\newblock {Nascent bipolar outflows associated with the first hydrostatic core
  candidates Barnard 1b-N and 1b-S}.
\newblock \emph{\aap} 577, L2.
\newblock \doi{10.1051/0004-6361/201525777}
\bibAnnoteFile{gerin2015}

\bibitem[{{Harsono} et~al.(2015){Harsono}, {van Dishoeck}, {Bruderer}, {Li},
  and {J{\o}rgensen}}]{harsono2015}
{Harsono}, D., {van Dishoeck}, E.~F., {Bruderer}, S., {Li}, Z.~Y., and
  {J{\o}rgensen}, J.~K. (2015).
\newblock {Testing protostellar disk formation models with ALMA observations}.
\newblock \emph{\aap} 577, A22.
\newblock \doi{10.1051/0004-6361/201424550}
\bibAnnoteFile{harsono2015}

\bibitem[{{Hatchell} et~al.(2007){Hatchell}, {Fuller}, {Richer}, {Harries}, and
  {Ladd}}]{hatchell2007}
{Hatchell}, J., {Fuller}, G.~A., {Richer}, J.~S., {Harries}, T.~J., and {Ladd},
  E.~F. (2007).
\newblock {Star formation in Perseus. II. SEDs, classification, and lifetimes}.
\newblock \emph{\aap} 468, 1009--1024.
\newblock \doi{10.1051/0004-6361:20066466}
\bibAnnoteFile{hatchell2007}

\bibitem[{{Hennebelle} and {Ciardi}(2009)}]{hennebelle2009aa}
{Hennebelle}, P. and {Ciardi}, A. (2009).
\newblock {Disk formation during collapse of magnetized protostellar cores}.
\newblock \emph{\aap} 506, L29--L32.
\newblock \doi{10.1051/0004-6361/200913008}
\bibAnnoteFile{hennebelle2009aa}

\bibitem[{{Hennebelle} and {Fromang}(2008)}]{hennebelle2008aa}
{Hennebelle}, P. and {Fromang}, S. (2008).
\newblock {Magnetic processes in a collapsing dense core. I. Accretion and
  ejection}.
\newblock \emph{\aap} 477, 9--24.
\newblock \doi{10.1051/0004-6361:20078309}
\bibAnnoteFile{hennebelle2008aa}

\bibitem[{{Hennebelle} and {Teyssier}(2008)}]{hennebelle2008b}
{Hennebelle}, P. and {Teyssier}, R. (2008).
\newblock {Magnetic processes in a collapsing dense core. II. Fragmentation. Is
  there a fragmentation crisis?}
\newblock \emph{\aap} 477, 25--34.
\newblock \doi{10.1051/0004-6361:20078310}
\bibAnnoteFile{hennebelle2008b}

\bibitem[{{Hincelin} et~al.(2016){Hincelin}, {Commer{\c{c}}on}, {Wakelam},
  {Hersant}, {Guilloteau}, and {Herbst}}]{hincelin2016}
{Hincelin}, U., {Commer{\c{c}}on}, B., {Wakelam}, V., {Hersant}, F.,
  {Guilloteau}, S., and {Herbst}, E. (2016).
\newblock {Chemical and Physical Characterization of Collapsing Low-mass
  Prestellar Dense Cores}.
\newblock \emph{\apj} 822, 12.
\newblock \doi{10.3847/0004-637X/822/1/12}
\bibAnnoteFile{hincelin2016}

\bibitem[{{Hirano} and {Liu}(2014)}]{hirano2014}
{Hirano}, N. and {Liu}, F.-c. (2014).
\newblock {Two Extreme Young Objects in Barnard 1-b}.
\newblock \emph{\apj} 789, 50.
\newblock \doi{10.1088/0004-637X/789/1/50}
\bibAnnoteFile{hirano2014}

\bibitem[{{Jones} and {Bate}(2018)}]{jones2018aa}
{Jones}, M.~O. and {Bate}, M.~R. (2018).
\newblock {Sink particle radiative feedback in smoothed particle hydrodynamics
  models of star formation}.
\newblock \emph{\mnras} 480, 2562--2577.
\newblock \doi{10.1093/mnras/sty1969}
\bibAnnoteFile{jones2018aa}

\bibitem[{{Karnath} et~al.(2020){Karnath}, {Megeath}, {Tobin}, {Stutz}, {Li},
  {Sheehan} et~al.}]{karnarth2020}
{Karnath}, N., {Megeath}, S.~T., {Tobin}, J.~J., {Stutz}, A., {Li}, Z.~Y.,
  {Sheehan}, P., et~al. (2020).
\newblock {Detection of Irregular, Submillimeter Opaque Structures in the Orion
  Molecular Clouds: Protostars within 10,000 yr of Formation?}
\newblock \emph{\apj} 890, 129.
\newblock \doi{10.3847/1538-4357/ab659e}
\bibAnnoteFile{karnarth2020}

\bibitem[{{Keto} et~al.(2015){Keto}, {Caselli}, and {Rawlings}}]{keto2015}
{Keto}, E., {Caselli}, P., and {Rawlings}, J. (2015).
\newblock {The dynamics of collapsing cores and star formation}.
\newblock \emph{\mnras} 446, 3731--3740.
\newblock \doi{10.1093/mnras/stu2247}
\bibAnnoteFile{keto2015}

\bibitem[{{Keto} and {Rybicki}(2010)}]{keto2010aa}
{Keto}, E. and {Rybicki}, G. (2010).
\newblock {Modeling Molecular Hyperfine Line Emission}.
\newblock \emph{\apj} 716, 1315--1322.
\newblock \doi{10.1088/0004-637X/716/2/1315}
\bibAnnoteFile{keto2010aa}

\bibitem[{{Koga} et~al.(2022){Koga}, {Kawasaki}, and {Machida}}]{koga2022}
{Koga}, S., {Kawasaki}, Y., and {Machida}, M.~N. (2022).
\newblock {Implementation of dust particles in three-dimensional
  magnetohydrodynamics simulation: dust dynamics in a collapsing cloud core}.
\newblock \emph{\mnras} 515, 6073--6092.
\newblock \doi{10.1093/mnras/stac2115}
\bibAnnoteFile{koga2022}

\bibitem[{{K{\"o}nyves} et~al.(2015){K{\"o}nyves}, {Andr{\'e}}, {Men'shchikov},
  {Palmeirim}, {Arzoumanian}, {Schneider} et~al.}]{Konyves:2015aa}
{K{\"o}nyves}, V., {Andr{\'e}}, P., {Men'shchikov}, A., {Palmeirim}, P.,
  {Arzoumanian}, D., {Schneider}, N., et~al. (2015).
\newblock {A census of dense cores in the Aquila cloud complex: SPIRE/PACS
  observations from the Herschel Gould Belt survey}.
\newblock \emph{\aap} 584, A91.
\newblock \doi{10.1051/0004-6361/201525861}
\bibAnnoteFile{Konyves:2015aa}

\bibitem[{{Krasnopolsky} et~al.(2011){Krasnopolsky}, {Li}, and
  {Shang}}]{krasnopolsky2011}
{Krasnopolsky}, R., {Li}, Z.-Y., and {Shang}, H. (2011).
\newblock {Disk Formation in Magnetized Clouds Enabled by the Hall Effect}.
\newblock \emph{\apj} 733, 54.
\newblock \doi{10.1088/0004-637X/733/1/54}
\bibAnnoteFile{krasnopolsky2011}

\bibitem[{{Larson}(1969)}]{larson1969}
{Larson}, R.~B. (1969).
\newblock {Numerical calculations of the dynamics of collapsing proto-star}.
\newblock \emph{\mnras} 145, 271.
\newblock \doi{10.1093/mnras/145.3.271}
\bibAnnoteFile{larson1969}

\bibitem[{{Lebreuilly} et~al.(2019){Lebreuilly}, {Commer{\c{c}}on}, and
  {Laibe}}]{lebreuilly2019}
{Lebreuilly}, U., {Commer{\c{c}}on}, B., and {Laibe}, G. (2019).
\newblock {Small dust grain dynamics on adaptive mesh refinement grids. I.
  Methods}.
\newblock \emph{\aap} 626, A96.
\newblock \doi{10.1051/0004-6361/201834147}
\bibAnnoteFile{lebreuilly2019}

\bibitem[{{Lebreuilly} et~al.(2020){Lebreuilly}, {Commer{\c{c}}on}, and
  {Laibe}}]{lebreuilly2020}
{Lebreuilly}, U., {Commer{\c{c}}on}, B., and {Laibe}, G. (2020).
\newblock {Protostellar collapse: the conditions to form dust-rich
  protoplanetary disks}.
\newblock \emph{\aap} 641, A112.
\newblock \doi{10.1051/0004-6361/202038174}
\bibAnnoteFile{lebreuilly2020}

\bibitem[{{Lebreuilly} et~al.(2023){Lebreuilly}, {Vallucci-Goy}, {Guillet},
  {Lombart}, and {Marchand}}]{lebreuilly2023}
{Lebreuilly}, U., {Vallucci-Goy}, V., {Guillet}, V., {Lombart}, M., and
  {Marchand}, P. (2023).
\newblock {Protostellar collapse simulations in spherical geometry with dust
  coagulation and fragmentation}.
\newblock \emph{\mnras} 518, 3326--3343.
\newblock \doi{10.1093/mnras/stac3220}
\bibAnnoteFile{lebreuilly2023}

\bibitem[{{Lee} et~al.(2023{\natexlab{a}}){Lee}, {Jhan}, and
  {Moraghan}}]{lee2023youngdisc}
{Lee}, C.-F., {Jhan}, K.-S., and {Moraghan}, A. (2023{\natexlab{a}}).
\newblock {First Detection of a Linear Structure in the Midplane of the Young
  HH 211 Protostellar Disk: A Spiral Arm?}
\newblock \emph{\apjl} 951, L2.
\newblock \doi{10.3847/2041-8213/acdbca}
\bibAnnoteFile{lee2023youngdisc}

\bibitem[{{Lee} et~al.(2023{\natexlab{b}}){Lee}, {Baek}, {Lee}, {Jeong}, {Kim},
  {Aikawa} et~al.}]{lee2023corino}
{Lee}, J.-E., {Baek}, G., {Lee}, S., {Jeong}, J.-H., {Kim}, C.-H., {Aikawa},
  Y., et~al. (2023{\natexlab{b}}).
\newblock {Complex Organic Molecules in a Very Young Hot Corino, HOPS 373SW}.
\newblock \emph{\apj} 956, 43.
\newblock \doi{10.3847/1538-4357/ace34b}
\bibAnnoteFile{lee2023corino}

\bibitem[{{Lee} et~al.(2004){Lee}, {Bergin}, and {Evans}}]{lee2004aa}
{Lee}, J.-E., {Bergin}, E.~A., and {Evans}, I., Neal~J. (2004).
\newblock {Evolution of Chemistry and Molecular Line Profiles during
  Protostellar Collapse}.
\newblock \emph{\apj} 617, 360--383.
\newblock \doi{10.1086/425153}
\bibAnnoteFile{lee2004aa}

\bibitem[{{Levermore} and {Pomraning}(1981)}]{levermore1981aa}
{Levermore}, C.~D. and {Pomraning}, G.~C. (1981).
\newblock {A flux-limited diffusion theory}.
\newblock \emph{\apj} 248, 321--334.
\newblock \doi{10.1086/159157}
\bibAnnoteFile{levermore1981aa}

\bibitem[{{Lewis} and {Bate}(2017)}]{lewis2017}
{Lewis}, B.~T. and {Bate}, M.~R. (2017).
\newblock {The dependence of protostar formation on the geometry and strength
  of the initial magnetic field}.
\newblock \emph{\mnras} 467, 3324--3337.
\newblock \doi{10.1093/mnras/stx271}
\bibAnnoteFile{lewis2017}

\bibitem[{{Lewis} et~al.(2015){Lewis}, {Bate}, and {Price}}]{lewis2015}
{Lewis}, B.~T., {Bate}, M.~R., and {Price}, D.~J. (2015).
\newblock {Smoothed particle magnetohydrodynamic simulations of protostellar
  outflows with misaligned magnetic field and rotation axes}.
\newblock \emph{\mnras} 451, 288--299.
\newblock \doi{10.1093/mnras/stv957}
\bibAnnoteFile{lewis2015}

\bibitem[{{Lombardi} et~al.(2015){Lombardi}, {McInally}, and
  {Faber}}]{lombardi2015}
{Lombardi}, J.~C., {McInally}, W.~G., and {Faber}, J.~A. (2015).
\newblock {An efficient radiative cooling approximation for use in hydrodynamic
  simulations}.
\newblock \emph{\mnras} 447, 25--35.
\newblock \doi{10.1093/mnras/stu2432}
\bibAnnoteFile{lombardi2015}

\bibitem[{{Machida} and {Basu}(2019)}]{machidabasu2019}
{Machida}, M.~N. and {Basu}, S. (2019).
\newblock {The First Two Thousand Years of Star Formation}.
\newblock \emph{\apj} 876, 149.
\newblock \doi{10.3847/1538-4357/ab18a7}
\bibAnnoteFile{machidabasu2019}

\bibitem[{{Machida} and {Hosokawa}(2020)}]{machida2020}
{Machida}, M.~N. and {Hosokawa}, T. (2020).
\newblock {Failed and delayed protostellar outflows with high-mass accretion
  rates}.
\newblock \emph{\mnras} 499, 4490--4514.
\newblock \doi{10.1093/mnras/staa3139}
\bibAnnoteFile{machida2020}

\bibitem[{{Machida} et~al.(2006{\natexlab{a}}){Machida}, {Inutsuka}, and
  {Matsumoto}}]{machida2006aa}
{Machida}, M.~N., {Inutsuka}, S.-i., and {Matsumoto}, T. (2006{\natexlab{a}}).
\newblock {Second Core Formation and High-Speed Jets: Resistive
  Magnetohydrodynamic Nested Grid Simulations}.
\newblock \emph{\apj} 647, L151--L154.
\newblock \doi{10.1086/507179}
\bibAnnoteFile{machida2006aa}

\bibitem[{{Machida} et~al.(2008){Machida}, {Inutsuka}, and
  {Matsumoto}}]{machida2008aa}
{Machida}, M.~N., {Inutsuka}, S.-i., and {Matsumoto}, T. (2008).
\newblock {High- and Low-Velocity Magnetized Outflows in the Star Formation
  Process in a Gravitationally Collapsing Cloud}.
\newblock \emph{\apj} 676, 1088--1108.
\newblock \doi{10.1086/528364}
\bibAnnoteFile{machida2008aa}

\bibitem[{{Machida} et~al.(2010){Machida}, {Inutsuka}, and
  {Matsumoto}}]{machida2010}
{Machida}, M.~N., {Inutsuka}, S.-i., and {Matsumoto}, T. (2010).
\newblock {Formation Process of the Circumstellar Disk: Long-term Simulations
  in the Main Accretion Phase of Star Formation}.
\newblock \emph{\apj} 724, 1006--1020.
\newblock \doi{10.1088/0004-637X/724/2/1006}
\bibAnnoteFile{machida2010}

\bibitem[{{Machida} and {Matsumoto}(2011)}]{machida2011discs}
{Machida}, M.~N. and {Matsumoto}, T. (2011).
\newblock {The origin and formation of the circumstellar disc}.
\newblock \emph{\mnras} 413, 2767--2784.
\newblock \doi{10.1111/j.1365-2966.2011.18349.x}
\bibAnnoteFile{machida2011discs}

\bibitem[{{Machida} et~al.(2005{\natexlab{a}}){Machida}, {Matsumoto}, {Hanawa},
  and {Tomisaka}}]{machida2005b}
{Machida}, M.~N., {Matsumoto}, T., {Hanawa}, T., and {Tomisaka}, K.
  (2005{\natexlab{a}}).
\newblock {Collapse and fragmentation of rotating magnetized clouds - II.
  Binary formation and fragmentation of first cores}.
\newblock \emph{\mnras} 362, 382--402.
\newblock \doi{10.1111/j.1365-2966.2005.09327.x}
\bibAnnoteFile{machida2005b}

\bibitem[{{Machida} et~al.(2006{\natexlab{b}}){Machida}, {Matsumoto}, {Hanawa},
  and {Tomisaka}}]{machida2006_1}
{Machida}, M.~N., {Matsumoto}, T., {Hanawa}, T., and {Tomisaka}, K.
  (2006{\natexlab{b}}).
\newblock {Evolution of Rotating Molecular Cloud Core with Oblique Magnetic
  Field}.
\newblock \emph{\apj} 645, 1227--1245.
\newblock \doi{10.1086/504423}
\bibAnnoteFile{machida2006_1}

\bibitem[{{Machida} et~al.(2005{\natexlab{b}}){Machida}, {Matsumoto},
  {Tomisaka}, and {Hanawa}}]{machida2005aa}
{Machida}, M.~N., {Matsumoto}, T., {Tomisaka}, K., and {Hanawa}, T.
  (2005{\natexlab{b}}).
\newblock {Collapse and fragmentation of rotating magnetized clouds - I.
  Magnetic flux-spin relation}.
\newblock \emph{\mnras} 362, 369--381.
\newblock \doi{10.1111/j.1365-2966.2005.09297.x}
\bibAnnoteFile{machida2005aa}

\bibitem[{{Masson} et~al.(2016){Masson}, {Chabrier}, {Hennebelle}, {Vaytet},
  and {Commer{\c{c}}on}}]{masson2016}
{Masson}, J., {Chabrier}, G., {Hennebelle}, P., {Vaytet}, N., and
  {Commer{\c{c}}on}, B. (2016).
\newblock {Ambipolar diffusion in low-mass star formation. I. General
  comparison with the ideal magnetohydrodynamic case}.
\newblock \emph{\aap} 587, A32.
\newblock \doi{10.1051/0004-6361/201526371}
\bibAnnoteFile{masson2016}

\bibitem[{{Masunaga} and {Inutsuka}(2000)}]{masunaga2000}
{Masunaga}, H. and {Inutsuka}, S.-i. (2000).
\newblock {A Radiation Hydrodynamic Model for Protostellar Collapse. II. The
  Second Collapse and the Birth of a Protostar}.
\newblock \emph{\apj} 531, 350--365.
\newblock \doi{10.1086/308439}
\bibAnnoteFile{masunaga2000}

\bibitem[{{Masunaga} et~al.(1998){Masunaga}, {Miyama}, and
  {Inutsuka}}]{masunaga1998}
{Masunaga}, H., {Miyama}, S.~M., and {Inutsuka}, S.-i. (1998).
\newblock {A Radiation Hydrodynamic Model for Protostellar Collapse. I. The
  First Collapse}.
\newblock \emph{\apj} 495, 346--369.
\newblock \doi{10.1086/305281}
\bibAnnoteFile{masunaga1998}

\bibitem[{{Matsumoto} and {Tomisaka}(2004)}]{matsumoto2004}
{Matsumoto}, T. and {Tomisaka}, K. (2004).
\newblock {Directions of Outflows, Disks, Magnetic Fields, and Rotation of
  Young Stellar Objects in Collapsing Molecular Cloud Cores}.
\newblock \emph{\apj} 616, 266--282.
\newblock \doi{10.1086/424897}
\bibAnnoteFile{matsumoto2004}

\bibitem[{{Maureira} et~al.(2020){Maureira}, {Arce}, {Dunham}, {Mardones},
  {Guzm{\'a}n}, {Pineda} et~al.}]{maureira2020}
{Maureira}, M.~J., {Arce}, H.~G., {Dunham}, M.~M., {Mardones}, D.,
  {Guzm{\'a}n}, A.~E., {Pineda}, J.~E., et~al. (2020).
\newblock {ALMA observations of envelopes around first hydrostatic core
  candidates}.
\newblock \emph{\mnras} 499, 4394--4417.
\newblock \doi{10.1093/mnras/staa2894}
\bibAnnoteFile{maureira2020}

\bibitem[{{Maury} et~al.(2022){Maury}, {Hennebelle}, and {Girart}}]{maury2022}
{Maury}, A., {Hennebelle}, P., and {Girart}, J.~M. (2022).
\newblock {Recent progress with observations and models to characterize the
  magnetic fields from star-forming cores to protostellar disks}.
\newblock \emph{Frontiers in Astronomy and Space Sciences} 9, 949223.
\newblock \doi{10.3389/fspas.2022.949223}
\bibAnnoteFile{maury2022}

\bibitem[{{Maury} et~al.(2011){Maury}, {Andr{\'e}}, {Men'shchikov},
  {K{\"o}nyves}, and {Bontemps}}]{Maury:2011aa}
{Maury}, A.~J., {Andr{\'e}}, P., {Men'shchikov}, A., {K{\"o}nyves}, V., and
  {Bontemps}, S. (2011).
\newblock {The formation of active protoclusters in the Aquila rift: a
  millimeter continuum view}.
\newblock \emph{\aap} 535, A77.
\newblock \doi{10.1051/0004-6361/201117132}
\bibAnnoteFile{Maury:2011aa}

\bibitem[{{Mellon} and {Li}(2008)}]{mellon2008aa}
{Mellon}, R.~R. and {Li}, Z.-Y. (2008).
\newblock {Magnetic Braking and Protostellar Disk Formation: The Ideal MHD
  Limit}.
\newblock \emph{\apj} 681, 1356--1376.
\newblock \doi{10.1086/587542}
\bibAnnoteFile{mellon2008aa}

\bibitem[{{Ohashi} et~al.(2014){Ohashi}, {Saigo}, {Aso}, {Aikawa}, {Koyamatsu},
  {Machida} et~al.}]{ohashi2014aa}
{Ohashi}, N., {Saigo}, K., {Aso}, Y., {Aikawa}, Y., {Koyamatsu}, S., {Machida},
  M.~N., et~al. (2014).
\newblock {Formation of a Keplerian Disk in the Infalling Envelope around L1527
  IRS: Transformation from Infalling Motions to Kepler Motions}.
\newblock \emph{\apj} 796, 131.
\newblock \doi{10.1088/0004-637X/796/2/131}
\bibAnnoteFile{ohashi2014aa}

\bibitem[{{Omukai}(2007)}]{omukai2007}
{Omukai}, K. (2007).
\newblock {Observational Characteristics of the First Protostellar Cores}.
\newblock \emph{\pasj} 59, 589--606.
\newblock \doi{10.1093/pasj/59.3.589}
\bibAnnoteFile{omukai2007}

\bibitem[{{Penston}(1969)}]{penston1969}
{Penston}, M.~V. (1969).
\newblock {Dynamics of self-gravitating gaseous spheres-III. Analytical results
  in the free-fall of isothermal cases}.
\newblock \emph{\mnras} 144, 425.
\newblock \doi{10.1093/mnras/144.4.425}
\bibAnnoteFile{penston1969}

\bibitem[{{Pineda} et~al.(2011){Pineda}, {Arce}, {Schnee}, {Goodman}, {Bourke},
  {Foster} et~al.}]{pineda2011aa}
{Pineda}, J.~E., {Arce}, H.~G., {Schnee}, S., {Goodman}, A.~A., {Bourke}, T.,
  {Foster}, J.~B., et~al. (2011).
\newblock {The Enigmatic Core L1451-mm: A First Hydrostatic Core? Or a Hidden
  VeLLO?}
\newblock \emph{\apj} 743, 201.
\newblock \doi{10.1088/0004-637X/743/2/201}
\bibAnnoteFile{pineda2011aa}

\bibitem[{{Price} et~al.(2003){Price}, {Pringle}, and {King}}]{price2003}
{Price}, D.~J., {Pringle}, J.~E., and {King}, A.~R. (2003).
\newblock {A comparison of the acceleration mechanisms in young stellar objects
  and active galactic nuclei jets}.
\newblock \emph{\mnras} 339, 1223--1236.
\newblock \doi{10.1046/j.1365-8711.2003.06278.x}
\bibAnnoteFile{price2003}

\bibitem[{{Price} et~al.(2012){Price}, {Tricco}, and {Bate}}]{price2012ab}
{Price}, D.~J., {Tricco}, T.~S., and {Bate}, M.~R. (2012).
\newblock {Collimated jets from the first core}.
\newblock \emph{\mnras} 423, L45--L49.
\newblock \doi{10.1111/j.1745-3933.2012.01254.x}
\bibAnnoteFile{price2012ab}

\bibitem[{{Priestley} et~al.(2018){Priestley}, {Viti}, and
  {Williams}}]{priestley2018ab}
{Priestley}, F.~D., {Viti}, S., and {Williams}, D.~A. (2018).
\newblock {An Efficient Method for Determining the Chemical Evolution of
  Gravitationally Collapsing Prestellar Cores}.
\newblock \emph{\aj} 156, 51.
\newblock \doi{10.3847/1538-3881/aac957}
\bibAnnoteFile{priestley2018ab}

\bibitem[{{Priestley} et~al.(2023){Priestley}, {Whitworth}, and
  {Fogerty}}]{priestley2023}
{Priestley}, F.~D., {Whitworth}, A.~P., and {Fogerty}, E. (2023).
\newblock {Differences in chemical evolution between isolated and embedded
  prestellar cores}.
\newblock \emph{\mnras} 518, 4839--4844.
\newblock \doi{10.1093/mnras/stac3444}
\bibAnnoteFile{priestley2023}

\bibitem[{{Priestley} et~al.(2019){Priestley}, {Wurster}, and
  {Viti}}]{priestley2019}
{Priestley}, F.~D., {Wurster}, J., and {Viti}, S. (2019).
\newblock {Ambipolar diffusion and the molecular abundances in pre-stellar
  cores}.
\newblock \emph{\mnras} 488, 2357--2364.
\newblock \doi{10.1093/mnras/stz1869}
\bibAnnoteFile{priestley2019}

\bibitem[{{Priestley} et~al.(2021){Priestley}, {Wurster}, and
  {Viti}}]{priestley2021}
{Priestley}, F.~D., {Wurster}, J., and {Viti}, S. (2021).
\newblock {Erratum: Ambipolar diffusion and the molecular abundances in
  pre-stellar cores}.
\newblock \emph{\mnras} 503, 2899--2901.
\newblock \doi{10.1093/mnras/stab702}
\bibAnnoteFile{priestley2021}

\bibitem[{{Rawlings} and {Yates}(2001)}]{rawlings2001aa}
{Rawlings}, J.~M.~C. and {Yates}, J.~A. (2001).
\newblock {Modelling line profiles in infalling cores}.
\newblock \emph{\mnras} 326, 1423--1430.
\newblock \doi{10.1111/j.1365-2966.2001.04674.x}
\bibAnnoteFile{rawlings2001aa}

\bibitem[{Saigo and Tomisaka(2006)}]{st2006}
Saigo, K. and Tomisaka, K. (2006).
\newblock Evolution of first cores in rotating molecular cores.
\newblock \emph{\apj} 645, 381
\bibAnnoteFile{st2006}

\bibitem[{{Saigo} and {Tomisaka}(2011)}]{saigo2011}
{Saigo}, K. and {Tomisaka}, K. (2011).
\newblock Spectrum energy distribution and submillimeter image of a rotating
  first core.
\newblock \emph{\apj} 728, 78.
\newblock \doi{10.1088/0004-637X/728/2/78}
\bibAnnoteFile{saigo2011}

\bibitem[{{Saigo} et~al.(2008){Saigo}, {Tomisaka}, and
  {Matsumoto}}]{saigo2008aa}
{Saigo}, K., {Tomisaka}, K., and {Matsumoto}, T. (2008).
\newblock {Evolution of First Cores and Formation of Stellar Cores in Rotating
  Molecular Cloud Cores}.
\newblock \emph{\apj} 674, 997--1014.
\newblock \doi{10.1086/523888}
\bibAnnoteFile{saigo2008aa}

\bibitem[{{Schnee} et~al.(2010){Schnee}, {Enoch}, {Johnstone}, {Culverhouse},
  {Leitch}, {Marrone} et~al.}]{schnee2010}
{Schnee}, S., {Enoch}, M., {Johnstone}, D., {Culverhouse}, T., {Leitch}, E.,
  {Marrone}, D.~P., et~al. (2010).
\newblock {An Observed Lack of Substructure in Starless Cores}.
\newblock \emph{\apj} 718, 306--313.
\newblock \doi{10.1088/0004-637X/718/1/306}
\bibAnnoteFile{schnee2010}

\bibitem[{{Sch{\"o}nke} and {Tscharnuter}(2011)}]{schonke2011}
{Sch{\"o}nke}, J. and {Tscharnuter}, W.~M. (2011).
\newblock {Protostellar collapse of rotating cloud cores. Covering the complete
  first accretion period of the stellar core}.
\newblock \emph{\aap} 526, A139.
\newblock \doi{10.1051/0004-6361/201015734}
\bibAnnoteFile{schonke2011}

\bibitem[{{Sharma} et~al.(2023){Sharma}, {J{\o}rgensen}, {Gavino}, {Ohashi},
  {Tobin}, {Lin} et~al.}]{sharma2023}
{Sharma}, R., {J{\o}rgensen}, J.~K., {Gavino}, S., {Ohashi}, N., {Tobin},
  J.~J., {Lin}, Z.-Y.~D., et~al. (2023).
\newblock {Early Planet Formation in Embedded Disks (eDisk). IX.
  High-resolution ALMA Observations of the Class 0 Protostar R CrA IRS5N and
  Its Surroundings}.
\newblock \emph{\apj} 954, 69.
\newblock \doi{10.3847/1538-4357/ace35c}
\bibAnnoteFile{sharma2023}

\bibitem[{{Shu}(1977)}]{shu1977}
{Shu}, F.~H. (1977).
\newblock {Self-similar collapse of isothermal spheres and star formation.}
\newblock \emph{\apj} 214, 488--497.
\newblock \doi{10.1086/155274}
\bibAnnoteFile{shu1977}

\bibitem[{{Spear} et~al.(2021){Spear}, {Maureira}, {Arce}, {Pineda}, {Dunham},
  {Caselli} et~al.}]{spear2021}
{Spear}, S., {Maureira}, M.~J., {Arce}, H.~G., {Pineda}, J.~E., {Dunham}, M.,
  {Caselli}, P., et~al. (2021).
\newblock {VLA and NOEMA Views of Bok Globule CB 17: The Starless Nature of a
  Proposed First Hydrostatic Core Candidate}.
\newblock \emph{\apj} 923, 231.
\newblock \doi{10.3847/1538-4357/ac3083}
\bibAnnoteFile{spear2021}

\bibitem[{{Stamatellos} et~al.(2007){Stamatellos}, {Whitworth}, {Bisbas}, and
  {Goodwin}}]{stamatellos2007}
{Stamatellos}, D., {Whitworth}, A.~P., {Bisbas}, T., and {Goodwin}, S. (2007).
\newblock {Radiative transfer and the energy equation in SPH simulations of
  star formation}.
\newblock \emph{\aap} 475, 37--49.
\newblock \doi{10.1051/0004-6361:20077373}
\bibAnnoteFile{stamatellos2007}

\bibitem[{{Stamer} and {Inutsuka}(2018)}]{stamer2018aa}
{Stamer}, T. and {Inutsuka}, S.-i. (2018).
\newblock {Radiation Hydrodynamics Simulations of Spherical Protostellar
  Collapse for Very Low Mass Objects}.
\newblock \emph{\apj} 869, 179.
\newblock \doi{10.3847/1538-4357/aaee81}
\bibAnnoteFile{stamer2018aa}

\bibitem[{{Stutz} et~al.(2013){Stutz}, {Tobin}, {Stanke}, {Megeath}, {Fischer},
  {Robitaille} et~al.}]{stutz2013aa}
{Stutz}, A.~M., {Tobin}, J.~J., {Stanke}, T., {Megeath}, S.~T., {Fischer},
  W.~J., {Robitaille}, T., et~al. (2013).
\newblock {A Herschel and APEX Census of the Reddest Sources in Orion:
  Searching for the Youngest Protostars}.
\newblock \emph{\apj} 767, 36.
\newblock \doi{10.1088/0004-637X/767/1/36}
\bibAnnoteFile{stutz2013aa}

\bibitem[{{Tan} et~al.(2014){Tan}, {Beltr{\'a}n}, {Caselli}, {Fontani},
  {Fuente}, {Krumholz} et~al.}]{tan2014}
{Tan}, J.~C., {Beltr{\'a}n}, M.~T., {Caselli}, P., {Fontani}, F., {Fuente}, A.,
  {Krumholz}, M.~R., et~al. (2014).
\newblock {Massive Star Formation}.
\newblock In \emph{Protostars and Planets VI}, eds. H.~{Beuther}, R.~S.
  {Klessen}, C.~P. {Dullemond}, and T.~{Henning}. 149--172.
\newblock \doi{10.2458/azu_uapress_9780816531240-ch007}
\bibAnnoteFile{tan2014}

\bibitem[{{Tassis} et~al.(2012){Tassis}, {Willacy}, {Yorke}, and
  {Turner}}]{tassis2012jul}
{Tassis}, K., {Willacy}, K., {Yorke}, H.~W., and {Turner}, N.~J. (2012).
\newblock {Non-equilibrium Chemistry of Dynamically Evolving Prestellar Cores.
  I. Basic Magnetic and Non-magnetic Models and Parameter Studies}.
\newblock \emph{\apj} 753, 29.
\newblock \doi{10.1088/0004-637X/753/1/29}
\bibAnnoteFile{tassis2012jul}

\bibitem[{{Tomida} et~al.(2010{\natexlab{a}}){Tomida}, {Machida}, {Saigo},
  {Tomisaka}, and {Matsumoto}}]{tomida2010dec}
{Tomida}, K., {Machida}, M.~N., {Saigo}, K., {Tomisaka}, K., and {Matsumoto},
  T. (2010{\natexlab{a}}).
\newblock Exposed long-lifetime first core: A new model of first cores based on
  radiation hydrodynamics.
\newblock \emph{The Astrophysical Journal} 725, L239--L244.
\newblock \doi{10.1088/2041-8205/725/2/L239}
\bibAnnoteFile{tomida2010dec}

\bibitem[{{Tomida} et~al.(2015){Tomida}, {Okuzumi}, and {Machida}}]{tomida2015}
{Tomida}, K., {Okuzumi}, S., and {Machida}, M.~N. (2015).
\newblock {Radiation Magnetohydrodynamic Simulations of Protostellar Collapse:
  Nonideal Magnetohydrodynamic Effects and Early Formation of Circumstellar
  Disks}.
\newblock \emph{\apj} 801, 117.
\newblock \doi{10.1088/0004-637X/801/2/117}
\bibAnnoteFile{tomida2015}

\bibitem[{{Tomida} et~al.(2013){Tomida}, {Tomisaka}, {Matsumoto}, {Hori},
  {Okuzumi}, {Machida} et~al.}]{tomida2013aa}
{Tomida}, K., {Tomisaka}, K., {Matsumoto}, T., {Hori}, Y., {Okuzumi}, S.,
  {Machida}, M.~N., et~al. (2013).
\newblock {Radiation Magnetohydrodynamic Simulations of Protostellar Collapse:
  Protostellar Core Formation}.
\newblock \emph{\apj} 763, 6.
\newblock \doi{10.1088/0004-637X/763/1/6}
\bibAnnoteFile{tomida2013aa}

\bibitem[{{Tomida} et~al.(2010{\natexlab{b}}){Tomida}, {Tomisaka}, {Matsumoto},
  {Ohsuga}, {Machida}, and {Saigo}}]{tomida2010may}
{Tomida}, K., {Tomisaka}, K., {Matsumoto}, T., {Ohsuga}, K., {Machida}, M.~N.,
  and {Saigo}, K. (2010{\natexlab{b}}).
\newblock {Radiation Magnetohydrodynamics Simulation of Proto-stellar Collapse:
  Two-component Molecular Outflow}.
\newblock \emph{\apjl} 714, L58--L63.
\newblock \doi{10.1088/2041-8205/714/1/L58}
\bibAnnoteFile{tomida2010may}

\bibitem[{{Tomisaka}(1998)}]{tomisaka1998}
{Tomisaka}, K. (1998).
\newblock {Collapse-Driven Outflow in Star-Forming Molecular Cores}.
\newblock \emph{\apjl} 502, L163--L167.
\newblock \doi{10.1086/311504}
\bibAnnoteFile{tomisaka1998}

\bibitem[{{Tomisaka}(2002)}]{tomisaka2002aa}
{Tomisaka}, K. (2002).
\newblock {Collapse of Rotating Magnetized Molecular Cloud Cores and Mass
  Outflows}.
\newblock \emph{\apj} 575, 306--326.
\newblock \doi{10.1086/341133}
\bibAnnoteFile{tomisaka2002aa}

\bibitem[{{Tomisaka} and {Tomida}(2011)}]{tomisaka2011}
{Tomisaka}, K. and {Tomida}, K. (2011).
\newblock {Observational Identification of First Cores: Non-LTE Radiative
  Transfer Simulation}.
\newblock \emph{\pasj} 63, 1151--1164.
\newblock \doi{10.1093/pasj/63.5.1151}
\bibAnnoteFile{tomisaka2011}

\bibitem[{{Tritsis} et~al.(2022){Tritsis}, {Federrath}, {Willacy}, and
  {Tassis}}]{2022tritsis}
{Tritsis}, A., {Federrath}, C., {Willacy}, K., and {Tassis}, K. (2022).
\newblock {Non-ideal magnetohydrodynamic simulations of subcritical pre-stellar
  cores with non-equilibrium chemistry}.
\newblock \emph{\mnras} 510, 4420--4435.
\newblock \doi{10.1093/mnras/stab3740}
\bibAnnoteFile{2022tritsis}

\bibitem[{{Tscharnuter}(1987)}]{tscharnuter1987}
{Tscharnuter}, W.~M. (1987).
\newblock {A collapse model of the turbulent presolar nebula}.
\newblock \emph{\aap} 188, 55--73
\bibAnnoteFile{tscharnuter1987}

\bibitem[{{Tscharnuter} and {Gail}(2007)}]{tscharnuter2007}
{Tscharnuter}, W.~M. and {Gail}, H.~P. (2007).
\newblock {2-D preplanetary accretion disks. I. Hydrodynamics, chemistry, and
  mixing processes}.
\newblock \emph{\aap} 463, 369--392.
\newblock \doi{10.1051/0004-6361:20065794}
\bibAnnoteFile{tscharnuter2007}

\bibitem[{{Tsukamoto} et~al.(2015){Tsukamoto}, {Iwasaki}, {Okuzumi}, {Machida},
  and {Inutsuka}}]{tsukamoto2015aa}
{Tsukamoto}, Y., {Iwasaki}, K., {Okuzumi}, S., {Machida}, M.~N., and
  {Inutsuka}, S. (2015).
\newblock {Effects of Ohmic and ambipolar diffusion on formation and evolution
  of first cores, protostars, and circumstellar discs}.
\newblock \emph{\mnras} 452, 278--288.
\newblock \doi{10.1093/mnras/stv1290}
\bibAnnoteFile{tsukamoto2015aa}

\bibitem[{{van Weeren} et~al.(2009){van Weeren}, {Brinch}, and
  {Hogerheijde}}]{van-weeren2009aa}
{van Weeren}, R.~J., {Brinch}, C., and {Hogerheijde}, M.~R. (2009).
\newblock {Modeling the chemical evolution of a collapsing prestellar core in
  two spatial dimensions}.
\newblock \emph{\aap} 497, 773--787.
\newblock \doi{10.1051/0004-6361/200811471}
\bibAnnoteFile{van-weeren2009aa}

\bibitem[{{Vaytet} et~al.(2012){Vaytet}, {Audit}, {Chabrier}, {Commer{\c c}on},
  and {Masson}}]{vaytet2012}
{Vaytet}, N., {Audit}, E., {Chabrier}, G., {Commer{\c c}on}, B., and {Masson},
  J. (2012).
\newblock {Simulations of protostellar collapse using multigroup radiation
  hydrodynamics. I. The first collapse}.
\newblock \emph{\aap} 543, A60.
\newblock \doi{10.1051/0004-6361/201219427}
\bibAnnoteFile{vaytet2012}

\bibitem[{{Vaytet} et~al.(2013){Vaytet}, {Chabrier}, {Audit}, {Commer{\c c}on},
  {Masson}, {Ferguson} et~al.}]{vaytet2013}
{Vaytet}, N., {Chabrier}, G., {Audit}, E., {Commer{\c c}on}, B., {Masson}, J.,
  {Ferguson}, J., et~al. (2013).
\newblock {Simulations of protostellar collapse using multigroup radiation
  hydrodynamics. II. The second collapse}.
\newblock \emph{\aap} 557, A90.
\newblock \doi{10.1051/0004-6361/201321423}
\bibAnnoteFile{vaytet2013}

\bibitem[{{Vaytet} and {Haugb{\o}lle}(2017)}]{vaytet2017}
{Vaytet}, N. and {Haugb{\o}lle}, T. (2017).
\newblock {A grid of one-dimensional low-mass star formation collapse models}.
\newblock \emph{\aap} 598, A116.
\newblock \doi{10.1051/0004-6361/201628194}
\bibAnnoteFile{vaytet2017}

\bibitem[{{Wakelam} et~al.(2022){Wakelam}, {Coutens}, {Gratier}, {Vidal}, and
  {Vaytet}}]{wakelam2022}
{Wakelam}, V., {Coutens}, A., {Gratier}, P., {Vidal}, T.~H.~G., and {Vaytet},
  N. (2022).
\newblock {Confirmation of the outflow in L1451-mm: SiO line and CH$_{3}$OH
  maser detections}.
\newblock \emph{\aap} 666, A191.
\newblock \doi{10.1051/0004-6361/202243459}
\bibAnnoteFile{wakelam2022}

\bibitem[{{Whitehouse} and {Bate}(2006)}]{whitehouse2006}
{Whitehouse}, S.~C. and {Bate}, M.~R. (2006).
\newblock The thermodynamics of collapsing molecular cloud cores using smoothed
  particle hydrodynamics with radiative transfer.
\newblock \emph{\mnras} 367, 32--38.
\newblock \doi{10.1111/j.1365-2966.2005.09950.x}
\bibAnnoteFile{whitehouse2006}

\bibitem[{{Whitworth} et~al.(1996){Whitworth}, {Bhattal}, {Francis}, and
  {Watkins}}]{whitworth1996}
{Whitworth}, A.~P., {Bhattal}, A.~S., {Francis}, N., and {Watkins}, S.~J.
  (1996).
\newblock {Star formation and the singular isothermal sphere}.
\newblock \emph{\mnras} 283, 1061--1070.
\newblock \doi{10.1093/mnras/283.3.1061}
\bibAnnoteFile{whitworth1996}

\bibitem[{{Wurster} et~al.(2021){Wurster}, {Bate}, and
  {Bonnell}}]{wurster2021b}
{Wurster}, J., {Bate}, M.~R., and {Bonnell}, I.~A. (2021).
\newblock {The impact of non-ideal magnetohydrodynamic processes on discs,
  outflows, counter-rotation, and magnetic walls during the early stages of
  star formation}.
\newblock \emph{\mnras} 507, 2354--2372.
\newblock \doi{10.1093/mnras/stab2296}
\bibAnnoteFile{wurster2021b}

\bibitem[{{Wurster} et~al.(2018){Wurster}, {Bate}, and {Price}}]{wurster2018c}
{Wurster}, J., {Bate}, M.~R., and {Price}, D.~J. (2018).
\newblock {Hall effect-driven formation of gravitationally unstable discs in
  magnetized molecular cloud cores}.
\newblock \emph{\mnras} 480, 4434--4442.
\newblock \doi{10.1093/mnras/sty2212}
\bibAnnoteFile{wurster2018c}

\bibitem[{{Yamada} et~al.(2009){Yamada}, {Machida}, {Inutsuka}, and
  {Tomisaka}}]{yamada2009}
{Yamada}, M., {Machida}, M.~N., {Inutsuka}, S.-i., and {Tomisaka}, K. (2009).
\newblock {Emission from a Young Protostellar Object. I. Signatures of Young
  Embedded Outflows}.
\newblock \emph{\apj} 703, 1141--1158.
\newblock \doi{10.1088/0004-637X/703/1/1141}
\bibAnnoteFile{yamada2009}

\bibitem[{{Yin} et~al.(2021){Yin}, {Priestley}, and {Wurster}}]{yin2021}
{Yin}, C., {Priestley}, F.~D., and {Wurster}, J. (2021).
\newblock {Investigating the role of magnetic fields in star formation using
  molecular line profiles}.
\newblock \emph{\mnras} 504, 2381--2389.
\newblock \doi{10.1093/mnras/stab1039}
\bibAnnoteFile{yin2021}

\bibitem[{{Yorke} et~al.(1993){Yorke}, {Bodenheimer}, and
  {Laughlin}}]{yorke1993}
{Yorke}, H.~W., {Bodenheimer}, P., and {Laughlin}, G. (1993).
\newblock {The Formation of Protostellar Disks. I. 1 M sub sun}.
\newblock \emph{\apj} 411, 274.
\newblock \doi{10.1086/172827}
\bibAnnoteFile{yorke1993}

\bibitem[{{Young}(2019)}]{young_thesis}
{Young}, A.~K. (2019).
\newblock \emph{Observational characteristics of early star formation}.
\newblock Phd thesis, University of Exeter, Exeter, U.K.
\newblock \doi{http://hdl.handle.net/10871/3942}
\bibAnnoteFile{young_thesis}

\bibitem[{{Young} et~al.(2019){Young}, {Bate}, {Harries}, and
  {Acreman}}]{young2019}
{Young}, A.~K., {Bate}, M.~R., {Harries}, T.~J., and {Acreman}, D.~M. (2019).
\newblock {Synthetic molecular line observations of the first hydrostatic core
  from chemical calculations}.
\newblock \emph{\mnras} 487, 2853--2873.
\newblock \doi{10.1093/mnras/stz1485}
\bibAnnoteFile{young2019}

\bibitem[{{Young} et~al.(2018){Young}, {Bate}, {Mowat}, {Hatchell}, and
  {Harries}}]{young2018aa}
{Young}, A.~K., {Bate}, M.~R., {Mowat}, C.~F., {Hatchell}, J., and {Harries},
  T.~J. (2018).
\newblock {What can the SEDs of first hydrostatic core candidates reveal about
  their nature?}
\newblock \emph{\mnras} 474, 800--823.
\newblock \doi{10.1093/mnras/stx2669}
\bibAnnoteFile{young2018aa}

\end{thebibliography}

\end{document}